\documentclass[apj]{emulateapj}

\usepackage{threeparttable}
\newcommand\B{\rule{.5ex}{0pt}}

\shorttitle{Giga-$z$: A 100,000 Object Superconducting Spectrophotometer for LSST Follow-up}
\shortauthors{D. W. Marsden {\it et al.}}

\begin{document}


\title{Giga-$z$: A 100,000 Object Superconducting Spectrophotometer for LSST Follow-up}  

\author{Danica~W.~Marsden and Benjamin~A.~Mazin} 
\affil{Department of Physics, University of California, Santa Barbara, CA 93106, USA}

\author{Kieran~O'Brien}
\affil{Department of Physics, University of Oxford, Denys Wilkinson Building, Keble Road, Oxford, OX1 3RH, UK}

\author{Chris~Hirata}
\affil{Department of Astrophysics, California Institute of Technology, 1216 East California Boulevard, Pasadena, CA 91106, USA}

\begin{abstract}     
We simulate the performance of a new type of instrument, a Superconducting Multi-Object Spectrograph (SuperMOS), that uses Microwave Kinetic Inductance Detectors (MKIDs).  MKIDs, a new detector technology, feature good QE in the UVOIR, can count individual photons with microsecond timing accuracy and, like X-ray calorimeters, determine their energy to several percent.  The performance of Giga-$z$, a SuperMOS designed for wide field imaging follow-up observations, is evaluated using simulated observations of the COSMOS mock catalog with an array of 100,000 R$_{423\,\rm nm}$\,=\,E/$\Delta$\,E\,=\,30 MKID pixels.  We compare our results against a simultaneous simulation of LSST observations.  In three years on a dedicated 4\,m-class telescope, Giga-$z$ could observe $\approx$\,2 billion galaxies, yielding a low resolution spectral energy distribution (SED) spanning 350\,-\,1350\,nm for each; 1000 times the number measured with any currently proposed LSST spectroscopic follow-up, at a fraction of the cost and time.  Giga-$z$ would provide redshifts for galaxies up to $z$\,$\approx$\,6 with magnitudes $m_i$\,$\lesssim$\,25, with accuracy $\sigma_{\Delta z /(1+z)}$\,$\approx$\,0.03 for the whole sample, and $\sigma_{\Delta z /(1+z)}$\,$\approx$\,0.007 for a select subset.  We also find catastrophic failure rates and biases that are consistently lower than for LSST.  The added constraint on Dark Energy parameters for WL\,+\,CMB by Giga-$z$ using the FoMSWG default model is equivalent to multiplying the LSST Fisher matrix by a factor of $\alpha$\,=\,1.27\,($w_p$), 1.53\,($w_a$), or 1.98\,($\Delta \gamma$).  This is equivalent to multiplying both the LSST coverage area and the training sets by $\alpha$, and reducing all systematics by a factor of 1/$\sqrt{\alpha}$, advantages that are robust to even more extreme models of intrinsic alignment. \end{abstract}



\section{Introduction}    

The accelerated expansion of the Universe \citep{perlmutter_1999, reiss_1998} is commonly attributed to a negative pressure component dubbed Dark Energy, making up approximately 73\% of the energy content of the Universe \citep{komatsu_2011}. The nature of Dark Energy remains a mystery, though it can be probed through its effect on the growth of structure over cosmic time.  As a result, the experiments aimed at understanding Dark Energy are quickly growing in number.  Galaxy surveys to map large scale structure and probe cosmology are becoming increasingly ambitious -- both in terms of the cosmological volumes they probe, as well as in the development of technological advances necessary for more precise and efficient measurement of galaxy spectral energy distributions (SEDs).  For example, the Large Synoptic Survey Telescope \citep[LSST;][]{lsst_2009} plans to image $\approx\,$10 billion galaxies to $m_i$\,$<$\,26.5 with data in the $u, g, r, i, z, y$ photometric bands over $\approx\,$20,000\,deg$^2$ of the Southern sky.  Similar current and future wide field imaging experiments include the Dark Energy Survey \citep[DES;][]{des_2005}, EUCLID \citep{euclid_2011}, and KIDS \citep{dejong_2012}.  Traditionally, sources selected by color and/or magnitude from initial imaging data in a handful of frequency bands were followed up with conventional dispersed spectrographs in order to obtain accurate redshifts.  However, in the coming data rich era, this approach is not possible.   Even the largest planned fiber-fed multi-object spectrographs cannot hope to follow up even 1\% of the LSST catalog \citep{schlegel_2011}.

One of the most important LSST science goals uses independent probes to measure the effect of Dark Energy: weak gravitational lensing (WL), Baryon Acoustic Oscillations (BAOs) and galaxy clusters \citep[][and references therein]{weinberg_2012}.  All of these techniques, however, rely on the precise determination of redshifts for as many galaxies and quasars as possible \citep{peacock_2006}, most of which will be faint given the steeply rising number counts towards fainter magnitudes \citep[e.g.,][]{smail_1995}.  Redshifts estimated from galaxy colors in a handful of broad bands have significant problems \citep{benitez_2009,hildebrandt_2010} since photometric accuracy depends on spectral coverage, resolution and signal-to-noise (S/N).  The biases and the high catastrophic failure rates that result from redshift determination using standard photometry add significant errors to the Dark Energy measurements \citep{wang_2010, bernstein_2010, hearin_2010}.  This naturally leads away from broad band imaging towards massively multiplexed low resolution spectroscopy or spectrophotometry. 
 
We consider a new instrument and survey, Giga-$z$, that will take low resolution spectra and find the redshifts of two billion objects in the LSST field down to $\lesssim$\,25$^{th}$ magnitude in $i$ band.  This survey, when combined with LSST imaging, will enable unique galaxy science.   Giga-$z$ is made possible by optical through near-IR Microwave Kinetic Inductance Detectors \citep[MKIDs;][]{day_2003}, a low temperature detector (LTD) developed at UCSB that can detect the energy and arrival time of each incoming photon without the use of bandpass filters or dispersive optics \citep{mazin_2012}.  MKIDs, described in Section~\ref{sec:mkids}, are nearly ideal, noiseless photon detectors, as they do not suffer from read noise or dark current, and have nearly perfect cosmic ray rejection.  In Giga-$z$, described in Section~\ref{sec:expt}, MKIDs will be used in a configuration similar to a conventional multi-object spectrograph, but without the use of a wavelength dispersive element.  Giga-$z$ could be on the sky by 2020, and with 3 years on a 4-m telescope could improve on the LSST constraints for $w$, the Dark Energy equation of state parameter, and $w_a$, its evolution, and in conjunction with LSST map the distribution of Dark Matter \citep[e.g.,][]{bacon_2005,kitching_2007}.  

The rest of this paper is organized as follows.  Section \ref{sec:sim_obs} explains the development of mock catalogs from simulated observations for both LSST and Giga-$z$.  Section \ref{sec:photozs} describes the redshift estimation, and compares results for the two experiments, as well as a summary of statistics for current or planned survey projects with similar science goals.  We explore Dark Energy parameter constraints in Section \ref{sec:de}, and conclude in Section \ref{sec:conc}.  


\section{Microwave Kinetic Inductance Detectors}      
\label{sec:mkids}

Large formats, high quantum efficiency, and low readout noise make semiconductor detectors the most popular type of detector used in the optical and near-IR wavelength regime.   However, thermal noise from their high ($\approx\,$100\,K) operating temperatures and the semiconductor band gap place fundamental limits.  Reducing gap parameters by a factor of a thousand can be achieved with cryogenic superconducting detectors, operating at around 100\,mK.  A superconducting detector can count single photons with no false counts while determining the energy (to a few percent) and arrival time (to roughly 1$\mu$s) of the incoming photon.   Since the photon energy is always much greater than the gap energy, much broader wavelength coverage is possible, enabling observations at infrared wavelengths that are vital to understanding the high redshift universe.  

A cryogenic detector technology with sensitivity and ease of multiplexing initially demonstrated at millimeter wavelengths \citep{roesch_2010, schlaerth_2010} are MKIDs \citep{day_2003}.  Intrinsic frequency domain multiplexing allows thousands of pixels to be read out over a single microwave cable  \citep{mchugh_2012}. They can count individual photons with no false counts and determine the energy and arrival time of every photon with good quantum efficiency \citep{mazin_2012}. Their physical pixel size and maximum count rate are well matched with large telescopes. These capabilities enable powerful new astrophysical instruments usable from the ground and space.  The MKIDs described here are sensitive to 0.1--5\,$\mu$m wavelength radiation (with cutoffs imposed by the sky count rate and the properties of the materials being used) but are optimized for near infrared (nIR) and optical wavelengths (350--1350\,nm).  

\begin{figure}[h!]
  \includegraphics[scale=0.33, clip, trim=0mm 0mm 0mm 0mm]{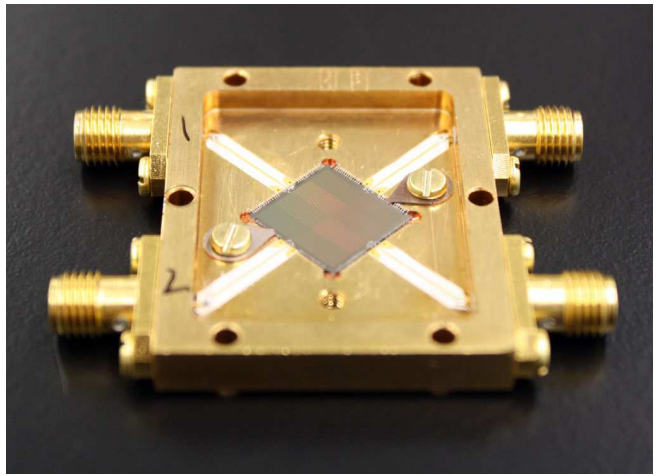}
  \caption{ A photograph of the new ARCONS 2,024 pixel MKID array mounted into a microwave package.  Signals are read out along 2 coaxial cables connected to the ports at each side of the box.  }
  \label{fig:mkids}
\end{figure}

The ARray Camera for Optical to Near-IR Spectrophotometry (ARCONS) is the first ever optical/nIR MKID camera.  It was commissioned in July, 2011 at the Palomar 200 inch telescope and as of December 2012, has now observed over a combined 24 nights on the Lick and Palomar telescopes \citep{mazin_2010, obrien_2012, mazin_2013}.  Some of the science targets observed include interacting binaries (AM Cvns, LMXBs, short period eclipsing sources), QSOs (for low resolution redshift measurements), supernovae (Type Ia and Type II) and the Crab pulsar.  ARCONS, representing the current state of optical MKIDs, houses a 2,024 detector array (Figure \ref{fig:mkids}), making it the largest optical/UV camera based on low temperature detectors by an order of magnitude.  

The energy resolution of the devices, R\,(=\,E/$\Delta$E), currently about 20 at 254\,nm (or about 12 at 423\,nm), can reasonably be expected to continue to improve towards the theoretical limit of 150 at 254\,nm over the next several years as designs and materials evolve.  Furthermore, the parallel technologies of infrared-blocking filters, broadband antireflection coatings, and detector quantum efficiency continue to develop, which will increase the performance of ARCONS and Giga-$z$.


\section{The Giga-$z$ Experiment}      
\label{sec:expt}   

Conventional multi-object spectrographs employ a mask inserted at the focal plane to pass light from targets through the slits (or apertures), blocking background sky and other nearby source photons to reduce sky noise and contamination.  A dispersive element such as a diffraction grating or prism then spreads the light as a function of wavelength on a detector.  

\begin{figure}[h!]
  \includegraphics[scale=0.6, clip, trim=0mm 0mm 0mm 0mm]{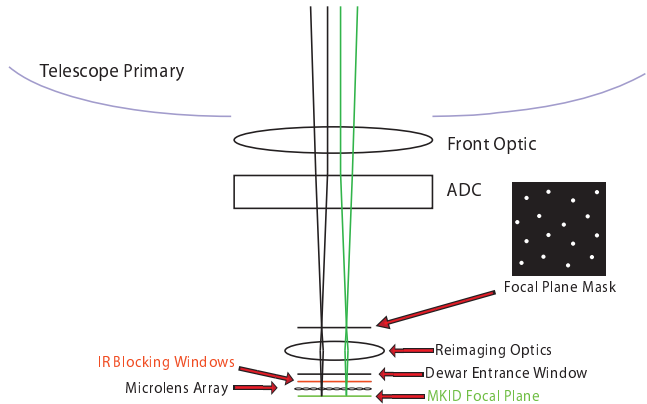}
  \caption{  After the secondary mirror, light passes through the primary mirror, and is corrected for atmospheric dispersion if required.  An aperture mask at the focal plane feeds preselected target light through a reimaging system which focuses the image onto the corresponding MKID.  Filters at 4\,K and 100\,mK block thermal infrared radiation.}
  \label{fig:optics}
\end{figure}

The SuperMOS concept uses the same mask-based approach to reduce sky background and contamination from other sources, but uses the intrinsic energy resolving capability of each MKID detector to measure the spectrum.  Since each MKID pixel provides spectral information the focal plane is used much more efficiently, yielding a simple and compact system.  A very simple implementation for Giga-$z$ is shown in Figure~\ref{fig:optics}, envisioned as an instrument for the Cassegrain or Naysmith focus of a dedicated 4\,m class telescope.

\begin{figure}[h!]
  \includegraphics[scale=0.9, clip, trim=0mm 0mm 0mm 0mm]{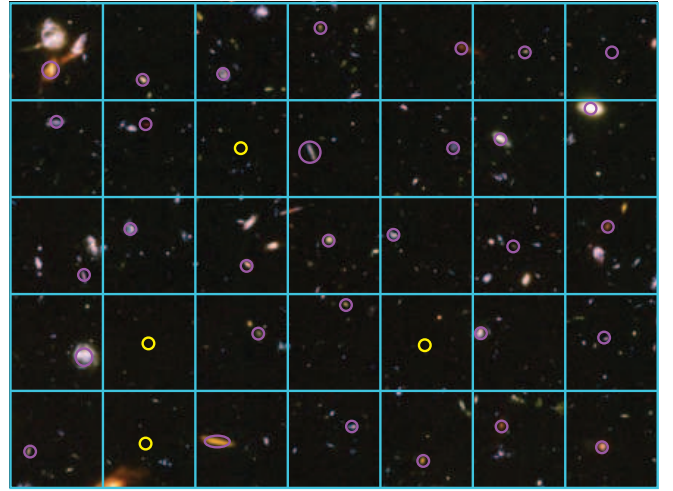}
  \caption{  A 1 deg$^2$ FOV is divided into 100,000 10''\,x\,10'' macropixels.   The background Hubble UDF image is for illustrative purposes and not to scale, with macropixels delineated in blue.  Existing catalogs will be used to select a target for each macropixel, and a corresponding hole drilled into a metal mask (purple circle), with the diameter allowed to vary depending on the object size and shape.  Source light passing through the mask will land on the corresponding MKID with the same plate scale as the macropixel.  Some subset of macropixels for each field will be selected to monitor the sky background (yellow circles). }
  \label{fig:pix}
\end{figure}

Aside from the inherent energy resolution of MKIDs, Giga-$z$ is enabled by the large pixel counts possible with MKIDs.  One square degree field of view can be divided among 100,000 detectors, each fed by a macropixel covering 10''$\times$10'' of the sky to be able to cover 20,000 square degrees in a reasonable amount of time (see Figure~\ref{fig:pix}).   Galaxy number counts in $I$ band to 24.5$^{th}$ magnitude \citep[e.g.,][]{capak_2007} ensure that $\gtrsim$\,80\% of the macropixels will contain a galaxy at each pointing.  A mask cut using pre-existing LSST (or earlier Dark Energy Survey) imaging would permit light from one celestial source per macropixel into a reimaging system that focused the light onto the corresponding large plate scale MKID located directly below.   

We note two potential drawbacks to this aperture masking technique: $\approx$\,20,000 precut masks are required, and it limits the galaxy sampling to a relatively uniform spacing, making observations of galaxy clusters more difficult.  However, a dedicated laser mask-milling facility can address the first issue \citep[such masks have been made for e.g.,][]{conti_2001, coil_2011}, and careful survey design incorporating fields with multiple visits can ameliorate the second.

In $\approx$ 15 minutes adequate S/N can be achieved to determine the redshift of galaxies with magnitudes $\lesssim$\,25 (Section \ref{sec:photozs}).  Assuming 80\% of the macropixels contain a source, Giga-$z$ would acquire $\approx$\,320,000 spectra per hour.  Rapid mask changes can be performed with pre-loaded cartridges, and the photon counting nature of the MKIDs allows the mask to be aligned using real time feedback from the science array.  In one night with 8 hours of observing, this equates to $\approx$ 2.5 million spectra per night.  At this rate, the entire LSST field could be covered in about three years.  

\subsection{Masks}      
\label{subsec:mask}

LSST will provide galaxy shapes and radii.  Assuming that Poisson statistics from the sky dominates errors, the mask hole radius that maximizes S/N is proportional to the encircled energy squared over the area of the hole, which depends upon the light profile of the target.  For an object with a Gaussian profile, for example, the S/N is maximized by capturing $\approx$\,72\% of a galaxy's light.  For more realistic profiles, it is slightly less than this.  The Cosmos Mock Catalog (CMC; Section \ref{sec:cmc}) indicates that the majority of galaxies out to high redshift have half light radii equal to less than half an arcsecond.  This scale translates to hole diameters of $\approx$\,40\,$\mu$m in the design presented here, well within the limits of current laser drilling technology.  Seeing conditions at the site may broaden galaxy profiles, and therefore must also be taken into account when determining hole size.  MKIDs have a maximum count rate which can be tuned to some degree during fabrication to fall within the range $\approx$\,1,000--10,000 counts/pixel/s.  The hole sizes for very bright or large galaxies would likely require accounting for the maximum allowable photon count rate.  

\subsection{Sky Subtraction}      
\label{subsec:skysub}

When working into the near-IR, sky subtraction becomes a dominant concern.  For Giga-$z$, concurrent with galaxy target selection from the LSST imaging will be the selection of known dark areas of the sky.  Approximately 10--20\% of the macropixels in a typical observation, greater than 10,000 MKIDs, will collect approximately 1,000 photons per second from the sky (based on the Gemini South model\footnote{www.gemini.edu/sciops/telescopes-and-sites/observing-condition-constraints}), with each photon individually time tagged to within a microsecond.  This sky background data can then be used to build up a map consisting of spectra as a function of time at every point on the array, facilitating the subtraction of the sky background to the Poisson limit over the entire spectral range of the detectors. In $J$ band, for example, with a sky brightness of roughly 16.6 magnitudes per square arcsecond, a 24.5$^{th}$ magnitude galaxy with about half of its light falling in 1 square arcsecond would have a contrast ratio of $\approx$\,5e-4.  Figure \ref{fig:sedswptsjy} shows this measured at a few sigma in a 15 minute exposure.     

\subsection{Instrument Response}      
\label{subsec:instr_resp}

Different mask hole positions within a macropixel illuminate the MKID very slightly differently, so both the throughput and quantum efficiency of each MKID will need to be calibrated as a function of mask hole position.  Stray light, for example, may be an important factor, involving cross-talk from one mask hole to another's corresponding detector.  These differences can be calibrated through laboratory and on-sky testing.  No fringing effects have been observed with ARCONS.


\section{Simulated Observations}
\label{sec:sim_obs}

\subsection{The Cosmos Mock Catalog}       
\label{sec:cmc}

The COSMOS mock catalog \citep[CMC; ][]{jouvel_2009} makes use of the latest survey data gathered through deep extragalactic surveys.  It was specifically designed to be used to forecast the yields of future Dark Energy surveys, by converting the observed properties of each COSMOS galaxy into simulated properties that can then be viewed using any instrument configuration.  Thus we combine the synthetic galaxy spectra from the CMC with our instrument throughput model to generate catalogs of simulated Giga-$z$ (and LSST) observations.  

The CMC is based on the Cosmic Evolution Survey (COSMOS) observations \citep{capak_2007} which cover approximately 2 square degrees with 30-band photometry from multiple instruments spanning X-ray through radio frequencies.  The subset used as input for the CMC is the photometric redshift catalog, covering a central 1.24 square degree patch fully covered by HST/ACS imaging and not masked, providing a sample of 538,000 objects down to $i^+$\,$<$\,26.5  \citep{ilbert_2009}.  

For each galaxy, AGN, or star, a best fit template is assigned based on the 30-band photometry.  The same template is used for the redshift fit, and comes from a composite library of elliptical, spiral and starburst galaxy templates and stellar templates used by the $Le Phare$ photo-$z$ code\footnote{www.lam.oamp.fr/arnouts/LE PHARE.html}.  The best fit templates are redshifted and scaled to be in the observer's frame, assuming a perfect instrument (perfect efficiency, delta function PSF, etc.).

Two possible downsides to using these observations are the potential bias due to faint AGN contribution, and the possible bias in the redshift distribution at $z$\,$\gtrsim$\,1.25, where the photo-$z$'s from the observed catalog become degraded.  To ensure a representative population, the CMC was compared with catalogs from the HST Ultra Deep Field, GOODS and VVDS-DEEP surveys for galaxy count, color, redshift and emission-line distributions, and found to be consistent. 

The CMC was updated in 2011 (version July 15, 2011) to improve the estimation of emission line fluxes, incorporated in the simulated galaxy SEDs, which we have used for the simulations presented here.  Table \ref{tab:cmc_models} shows the breakdown by galaxy type of the CMC.  The newest version of the catalog contains 646,706 objects, to a limiting simulated Subaru $i$ band AB magnitude of 26.5 over 1.24 square degrees of sky.  This catalog has already had most stars and AGN removed.  For our simulations, we use CMC objects with $i$\,$<$\,25 (a complete sample), and remove the 916 AGN and 1373 point-like objects with no radius solution (likely stars; there are 2 objects which overlap the AGN/point-like object designations), which results in 219,759 galaxies for simulating mock observations, with redshifts up to $\approx$\,6.  

{
\renewcommand{\arraystretch}{1.3}

\begin{table}[h!]
\begin{center}
\caption{Distribution of the CMC by Galaxy type. \label{tab:cmc_models}}
\begin{tabular}{lccc}
\tableline
\tableline
  Type  &  N$_{gal}$ & $m_i\,<$\,25 & Non-AGN  \\
\tableline
Ell - S0      & 14927 & 11060 &  10985 \\
Sa - Sc      & 38246 & 7651 &   7471 \\
Sd - Sdm  & 35704 & 10290 &  10045 \\
Starburst   & 557829 & 193045 & 191258  \\
\tableline
Total          & 646706 & 222046 & 219759  \\
\tableline
\end{tabular}
\end{center}
\end{table}
}


\subsection{Simulated Giga-$z$ Observations}      
\label{sec:gigazobs}

To simulate the performance of Giga-$z$, realistic models for filters, optical throughput, device quantum efficiency (QE), telescope reflectivity and sky background were generated.  A locale with conditions similar to the Cerro Pach\'{o}n, Chile, with access to the southern sky is assumed.

We take optical throughput to be 0.7, accounting for $\approx$\,4\% loss at each of five lenses and $\approx$\,10\% loss at an IR-blocking filter.   Since the QE for the MKID detectors in ARCONS varied between roughly 73\% at 200\,nm and 22\% at 3\,$\mu$m \citep{mazin_2010}, and there is significant room for improvement (Section \ref{sec:mkids}), we assume a constant MKID QE of 0.75 for Giga-$z$.  A model for bare Aluminum reflectivity\footnote{http://rmico.com/coatings-specifications/metal-hybrid/bare-aluminum-bal} squared to account for two reflections at the primary and secondary mirrors is used to account for losses at reflective surfaces. 

To model the atmospheric transmission at Cerro Pach\'{o}n, Chile, we combine the extinction curve taken from the Gemini website\footnote{http://www.gemini.edu/sciops/telescopes-and-sites/observing -condition-constraints/extinction} in the optical, merged with transmission in the nIR\footnote{http://www.gemini.edu/?q=node/10789}.  To model the sky background for simulating measurement errors and estimates of S/N, we merge the optical and near infrared sky backgrounds given for Gemini South\footnote{http://www.gemini.edu/sciops/telescopes-and-sites/observing -condition-constraints/ir-background-spectra}$^,$\footnote{http://www.gemini.edu/sciops/telescopes-and-sites/observing -condition-constraints/optical-sky-background} assuming an airmass of 1.5 and water vapour column of 4.3\,mm.  This is likely an overestimate of the far-red continuum brightness \citep[e.g.,][]{hanuschik_2003}, but a subdominant effect at R$_{423}$\,=\,30, as OH lines will be the primary contributor and thus we deem this a conservative estimate.  The lunar phase assumed is within 7 nights from new moon (``dark'', but not ``darkest''), where the $V$ magnitude of the sky is $\approx$\,20.7\,mag/arcsec$^2$.  In practice, grey time can likely be used without significant degradation. 

MKIDs do not require the use of filters for information about the energy of the incoming photons.   However, in order to compare our results with the LSST simulated photometry as transparently as possible and to be able to use existing redshift estimation codes without having to develop our own at this time, we can simulate effective photometric bands as independent spectral resolution elements.  Collected photons are separated into energy bins, thus an effective filter would take the form of the error distribution of the energy determination convolved with a tophat, which represents the quantization of binning.  Since photon energies are determined by a fit to the phase shift of the MKID with time, we can choose bins small compared to the true resolution so as not to impose any additional degradation.  For the analysis presented here, each ``filter'' is a Gaussian, centered in wavelength one full width half maximum (FWHM) away from its neighbors, where 

\begin{equation}
\mathrm{FWHM} (\lambda) = 2\,\sqrt{2 \, ln (2)}\,\sigma(\lambda) = \frac{\lambda^2}{\mathrm{R}_0 \, \lambda_0} .
\end{equation}  

\begin{figure}[h!]
  \includegraphics[scale=0.51, clip, trim=5mm 3mm 0mm 1mm]{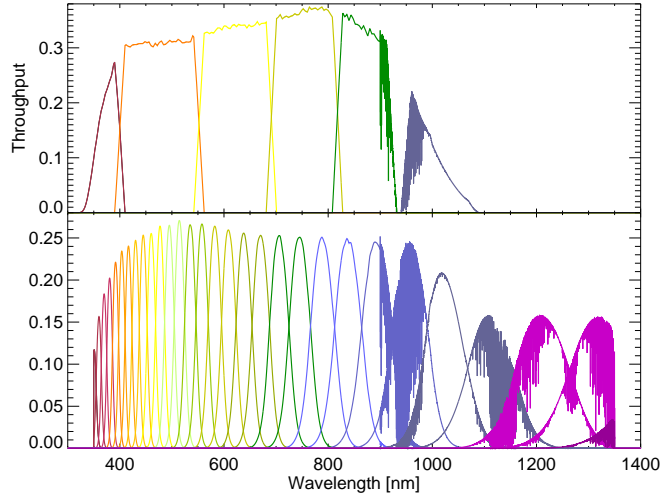}
  \caption{ The simulated system throughputs for the LSST (top) and Giga-$z$ (bottom) experiments, accounting for optical element (filter and lens) throughputs, mirror reflectivity, detector quantum efficiency and atmospheric transmission.  The LSST $u, g, r, i, z, y$ filter implementation (see Section \ref{sec:lsst}) is based on the \cite{lsst_2009} publication.  The effective Giga-$z$ filter set is for devices with energy resolution R\,=\,30 at 423\,nm.  Cutoffs are imposed at 350 and 1350\,nm due to the effects of sky brightness and atmospheric transmission constraints.}
  \label{fig:through}
\end{figure}

\noindent R$_0$ is equal to 30 at the fiducial $\lambda_0$ which we take here to be 423\,nm, and decreases linearly with increasing wavelengths.  Figure \ref{fig:through} (bottom panel) illustrates the effective filter set in the context of total throughput, using a normalization that ensures no double-counting of photons.  Cutoffs are imposed at  350 and 1350\,nm, where the sky brightness and atmospheric transmission present practical limits. 
 
It is important to note that although we distinguish frequency ``filters'' here, all wavelengths are observed simultaneously by Giga-$z$, resulting in extremely efficient use of exposure time.  As well, each pseudo-filter sees the same observing conditions with time, simplifying analysis considerably, a second major advantage over usual multi-filter photometry.  In practice, for an MKID array, a more optimal solution would be to take the photon events and construct a maximum likelihood algorithm to reconstruct the spectrum.

Combining these throughput models as a function of wavelength for the various loss mechanisms gives the total expected system throughput depicted in Figure \ref{fig:through} (bottom panel).   For the final simulated observations, noise was added to the observed fluxes according to the properties of our system.  

\subsection{Simulated LSST Observations}      
\label{sec:lsst}

\begin{figure}[h!]
  \includegraphics[scale=0.51, clip, trim=7mm 3mm 0mm 2mm]{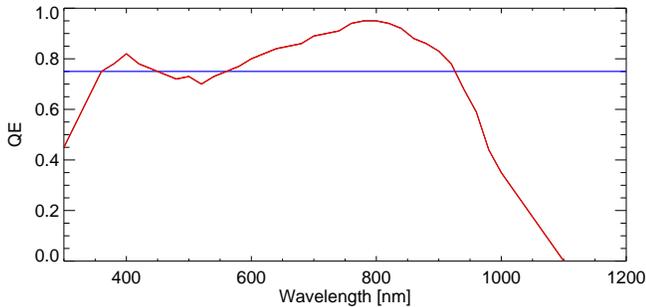}
  \caption[.]{ The QE for the red-enhanced devices developed at MIT/LL which are leading LSST device technology candidates \citep{lesser_2002} is shown in red.  The blue line depicts the assumed Giga-$z$ QE.}
  \label{fig:lsst_qe}
\end{figure}

We also simulated observations of the CMC for an LSST 3-year stack.  The LSST design, as outlined in the LSST Science book \cite{lsst_2009}, incorporates 3 lenses, each with a projected $\approx$\,3\% loss, giving a throughput of $\approx$\,91\%.  They use a 3 mirror telescope, which we model using the same aluminum reflectivity model but cubed.  We note that this is pessimistic compared with the proposed LSST design, that uses a multilayer mirror coating to improve on the far-red performance of the aluminum, without the $u$ band absorption of silver\footnote{http://lsst.org/files/docs/LSST-RefDesign.pdf}.  \cite{lesser_2002} predict that the devices used in the LSST experiment will be very similar to the red wavelength-enhanced charge coupled devices (CCDs) developed at the Massachusetts Institute of Technology (MIT) Lincoln Laboratory, with a QE shown in Figure \ref{fig:lsst_qe}.  Table \ref{tab:lsst_filts} lists the LSST filter band centers and bandwidth, as well as the exposure time for each filter.  In the LSST case, only one filter may be used at a time.

{
\renewcommand{\arraystretch}{1.3}

\begin{table}[h!]
\begin{center}
\caption{LSST Filters and Exposure Times. \label{tab:lsst_filts}}
\begin{tabular}{cccc}
\tableline
\tableline
  Filter  &  Central $\lambda$ & Bandwidth & Exposure Time  \\
   & [nm] & [nm] & [s] \\
\tableline
$u$      & 360.0 & 80.0 & 700  \\
$g$      & 476.0 & 152.0 & 1000 \\
$r$       & 621.0 & 69.0 & 2300 \\
$i$       & 754.5 & 63.5 & 2300 \\
$z$      & 870.0 & 52.0 & 2300 \\
$y$      & 1015.0 & 65.0 & 2000 \\
\tableline
\end{tabular}
\end{center}
\end{table}
}

\subsection{Simulation Output}    

\begin{figure}[h!]
  \includegraphics[scale=0.51, clip, trim=8mm 3mm 0mm 8mm]{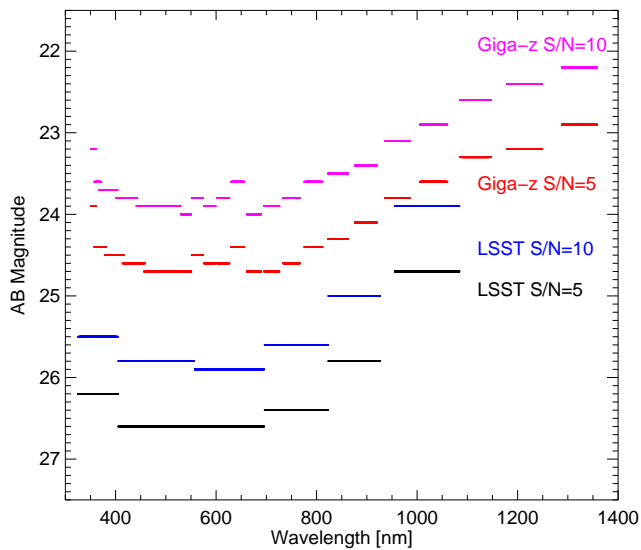}
  \caption{ The simulated 5$\sigma$ and 10$\sigma$ magnitude limits for a 3 year LSST stack and Giga-$z$ experiment. }
  \label{fig:ptsrc}
\end{figure}

Figure~\ref{fig:ptsrc} illustrates our simulated 5$\sigma$ and 10$\sigma$ magnitude limits for the LSST 3 year stack and Giga-$z$ experiment, accounting for optical element (filter and lens) throughputs, mirror reflectivity, detector quantum efficiency and atmospheric transmission.  Note that each LSST filter encompasses 2--5 Giga-$z$ pseudo-filters.  If the Giga-$z$ filters corresponding to each LSST filter were combined, the resulting magnitude limits would be more equivalent.
 
\begin{figure*}[ht!]
  \begin{center}
  \includegraphics[scale=.91, clip, trim=0mm 1mm 0mm 4mm]{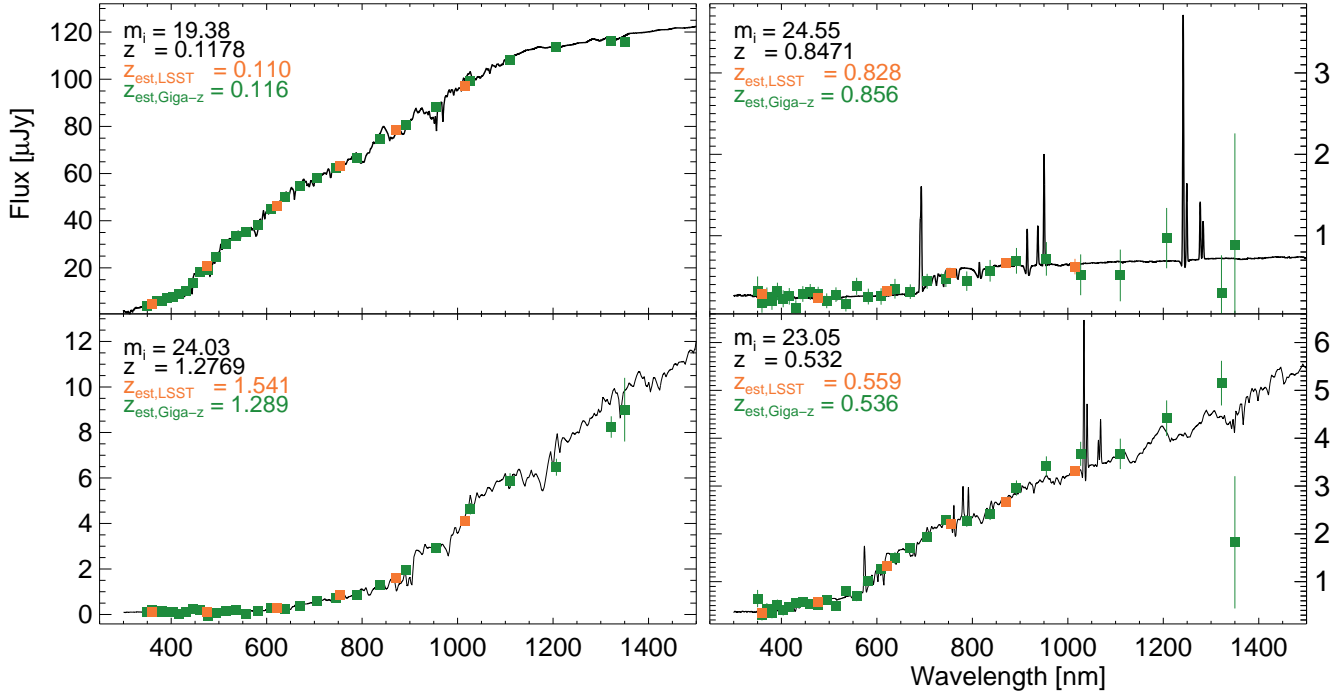}
  \end{center}
  \caption{ Examples of simulated measurements of CMC SEDs (black) for several different galaxy types in units of $\mu$Jy by LSST (3 year stack; orange) and Giga-$z$ (green), with Poisson errorbars.}
  \label{fig:sedswptsjy}
\end{figure*}

We show the simulated photometric measurements by LSST and Giga-$z$ for four example CMC mock galaxy spectra in Figure~\ref{fig:sedswptsjy}.  Orange points denote the LSST mock observations with errorbars, and green points those predicted for the Giga-$z$ experiment.  Errors are simply derived from Poisson statistics, and optimal sky background subtraction has been assumed.  Though the S/N is typically lower per filter for Giga-$z$, the wavelength coverage is greater, and the exposure time is much smaller for Giga-$z$.  Indeed, as will be seen in Section \ref{sec:photoz_res}, photometric depth does not equate to photometric redshift accuracy.  Despite high S/N, the LSST filter set (R\,$\approx$\,5) inevitably leads to more color-redshift degeneracies, making unambiguous redshift determination impossible for the majority of galaxies \citep[e.g.,][]{coe_2006}. 

\section{Photometric Redshifts}       
\label{sec:photozs}

Very accurate photo-$z$'s are required to optimally exploit the expected data sets of Dark Energy surveys such as DES, LSST and EUCLID \citep[e.g.,][]{huterer_2006}.  However, photo-$z$ accuracy is greatly affected by experimental observing strategy, filter set and photometric sensitivity, which will impact the determination of redshifts from broadband SEDs.  

The best photometric results arise when strong continuum breaks in a galaxy spectrum fall between two instrument filters, and are therefore well constrained.  One such typical feature of early-type galaxies is the 4000\,\AA\,break which arises from the onset of stellar photospheric opacity due to the absorption of mainly ionized metals (e.g. Ca\,II) in the atmospheres of late-type stars.  The Balmer break at 3646\,\AA\ marks the termination of the hydrogen Balmer series, and is indicative of younger stellar populations and more recent star-formation. The Lyman break is another pronounced continuum discontinuity at 912\,\AA, observed in star-forming galaxy spectra.  It is produced both in the stellar atmospheres of massive stars as a result of the hydrogen ionization edge and by photoelectric absorption by interstellar and intergalactic H\,I gas.   Lastly, sources at high redshifts have spectra that exhibit a suppressed continuum blueward of 1216\,\AA\ (Ly-$\alpha$) due to additional opacity from line blanketing by intervening gas clouds along the line of sight.

By $z$\,$\approx$\,1.25, the 4000\,\AA\ break has shifted to $\approx$\,0.9\,$\mu$m, out of the range of typical optical filters such as those used by LSST.  Not until $z$\,$\gtrsim$\,3, does the redshifted Lyman break re-enter the visible, so nIR or UV data becomes imperative to get reliable redshifts in this ``redshift desert''.    The wavelength coverage of Giga-$z$ is continuous and dense from $\approx$\,350\,nm--1.35\,$\mu$m (Figure \ref{fig:through}), narrowing the redshift desert to 2.25\,$\lesssim$\,$z$\,$\lesssim$\,3.  We therefore expect fewer photo-$z$ degeneracies and a lower catastrophic failure rate for Giga-$z$ than for LSST.


\subsection{Choosing a Photo-$z$ Code}       
\label{sec:whichcode}

\cite{hildebrandt_2010} compared 17 photo-$z$ estimation methods used currently in the literature through blind tests on both simulated data and real data from the Great Observatories Origins Deep Survey \citep[GOODS;][]{giavalisco_2004}.  Using each code, accuracies were determined for global photo-$z$ bias, scatter, and outlier rates.  Differences between codes stemmed mainly from whether they were empirical or template-fitting, the training set in the former case and the template set in the latter, the use of priors, handling of the Lyman-$\alpha$ forest and the benefit of adding mid-IR photometry.  For a detailed discussion of photo-$z$ methods and their common elements see \cite{budavari_2009}.   The three photo-$z$ codes which ranked highest were $Le Phare$ \citep{arnouts_2002, ilbert_2006}, Bayesian Photo-$z$'s \citep{benitez_2000, coe_2006} and EAZY \citep{brammer_2008}.  The best results were achieved by using an empirical code (smaller biases) or optimizing templates, and correcting for systematic offsets.  

One reason to favor a template-based photo-$z$ code without any required training for this study is that the galaxies probed by upcoming surveys will span a cosmological volume and parameter space much greater than what can be well represented currently through spectroscopy. Even very small mismatches between the mean photometric target and the training set can induce photo-$z$ biases large enough to corrupt derived cosmological parameters significantly \citep{macdonald_2010}.  Hence we proceed with the template-based photo-$z$ code EAZY, and interpret the results presented here as what could be achieved prior to comprehensive spectroscopic surveys.  Its ease of use and demonstrated improvement in photo-$z$'s with the inclusion of IR data are also factors in why EAZY is commonly used \citep[e.g.,][]{ly_2011}, facilitating our aim of a side-by-side comparison between LSST and Giga-$z$.  

Secondly, emission lines can change the colors of objects significantly, and are present in real observations.  The treatment of emission lines can improve the photo-$z$ accuracy by a factor of $\approx$\,2.5 \citep{ilbert_2009}, as they can be critical for minimizing systematic errors (such as aliasing in the redshifts).  Both the CMC (Section \ref{sec:cmc}) and templates provided with EAZY include emission lines.  

Lastly, we note that the CMC synthetic spectra from which our mock observations are derived were generated using the $Le Phare$ code and therefore it is more instructive to use a different code for our analysis.


\subsection{The EAZY Photometric Redshift Code}       
\label{sec:eazy}

EAZY\footnote{Easy and Accurate Redshifts from Yale; http://www.astro.yale.edu/eazy/} was developed to be specifically optimized for samples of galaxies with a limited amount of or biased (e.g., band-selected) spectroscopic information available \citep{brammer_2008}.  Combining the functionality of several pre-existing redshift estimation codes, EAZY allows the user to fit linear combinations of templates and the choice of using a prior (flux- and redshift-based) through a parameter file.   An error function can be used to downweight spectra to account for wavelength-dependent template mismatch such as in the UV where dust extinction is strongest and most variable, and in the nIR where thermal dust emission and stochastic PAH line features begin to appear.  Furthermore, the user may apply \cite{madau_1995} IGM absorption to templates.

By default, a probability-weighted integral is taken over the full redshift grid in order to assign an object redshift, marginalizing over the posterior redshift probability distribution (in lieu of e.g. assigning the single most likely redshift by $\chi^2$ minimization, although the user has control over which).  Though this does not permit simple spectral classification, the increased photo-$z$ precision and ability to reproduce complex star-formation histories by fitting non-negative linear combinations of the templates may allow for better physical separation of photometric samples.  One particular feature is the applicability to a wider range of redshifts and intrinsic colors than would be possible with an empirical photo-$z$ code, as no representative training set exists. 

\begin{figure}[h!]
  \includegraphics[scale=0.52, clip, trim=3mm 4mm 0mm 3mm]{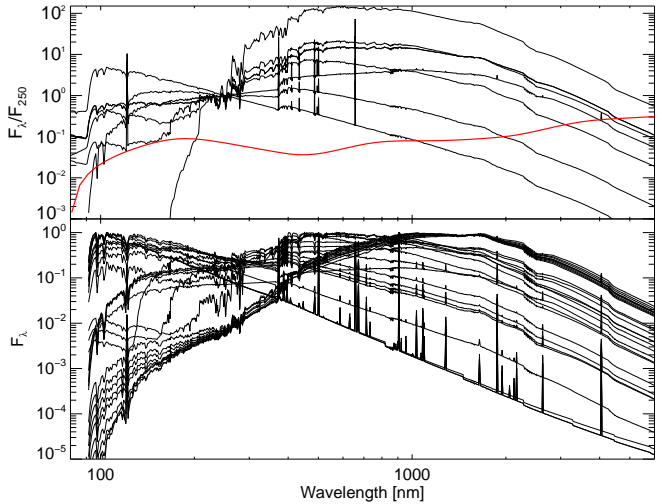}
  \caption{ EAZY template spectra (top panel) with the default error function plotted in red, and 26 of the 260 Pegase13 template spectra (bottom panel) used for redshift estimation with the EAZY code.  Details on the templates can be found in the text.}
  \label{fig:templates}
\end{figure}

The default EAZY template set was generated using the \cite{blanton_2007} non-negative matrix factorization method to reduce the stellar population synthesis code P\'{E}GASE model library \citep{fioc_1997} to five ``principle component'' spectral templates of the calibration catalog.  The templates are calibrated with a catalog \citep{blaizot_2005} derived from the \cite{delucia_2007} semi-analytic models based on the \cite{springel_2005} Millennium Simulation.  Theoretically, this simulated 1 deg$^2$ light cone contains a more realistic distribution of galaxies over 0\,$<$\,$z$\,$\lesssim$\,4 than the more local spectroscopy most codes are trained on. One additional dusty starburst template was added, however, to account for extremely dusty galaxies which appear to be lacking representation in the semi-analytic models.  A newer version of EAZY (v1.1; private communication with G. Brammer) implements emission lines following the prescription of \cite{ilbert_2009} \citep[after][]{kennicutt_1998}.  These template spectra are shown in the top panel of Figure \ref{fig:templates}. 

A second set of templates provided with EAZY that were used in this analysis are a grid of single P\'{E}GASE models (which we will refer to as ``Pegase13'') that provide a self-consistent treatment of emission lines. They were designed to match the set described by \cite{grazian_2006} - constant star formation rate models with additional dust reddening following the \cite{calzetti_2001} law.  One tenth of the 260 Pegase13 spectral templates are plotted in the bottom panel of Figure \ref{fig:templates}. 

The redshift grid for template fitting was done in steps of 0.005 on a log(1+$z$) scale.  This imposes a limit on redshift resolution that becomes important on scales of $| \Delta z |$/(1+$z$)\,$<$\,0.01, as features can be recovered if their wavelength is greater than two grid steps.  As our total sample redshift accuracy is larger than this, our sampling step is not the dominant contributor to systematic errors.  


\subsection{Photo-z Results}            
\label{sec:photoz_res}

In order to make the fairest comparison possible, we did not employ techniques which might aid in decreasing scatter in photo-$z$ estimates in a biased way.  For example, we perform simple cuts, but we do not use priors in this analysis as they may improve results in a way that is preferential for one experiment.  To minimize obscuration, we present only basic first order results that could be improved upon in the future when the goal is to get the most out of data.

The EAZY estimated redshifts for both Giga-$z$ and LSST were compared with the input redshifts from the CMC catalog, and the statistics used to quantitatively assess their quality are given in Table \ref{tab:zstats}.  The difference between input catalog and EAZY estimated redshift is $\Delta z$\,=\,$z_{in}\,$-$\,z_{est}$.  The distribution for $\Delta z$ is typically non-Gaussian, with extended tails and secondary peaks due to catastrophic outliers, which we arbitrarily define here as objects for which  $| \Delta z | /(1+z_{in})$\,$>$\,0.15. We quantify bias as $median[\Delta z/(1+z_{in})$], where the factor (1+$z_{in}$) accounts for the scaling of redshift errors with the stretching of rest-frame spectral features. 
    
 One common method for estimating the redshift accuracy from $\sigma_{\Delta z/(1+z_{in})}$ uses the normalized median absolute deviation \citep[NMAD:][after \cite{brammer_2008}]{hoaglin_1983}, defined here as
 \begin{equation}
\sigma_{\textrm{\tiny NMAD}} = 1.48 \times median \left( \left|  \frac{\Delta z - median(\Delta z) }{(1 + z_{in})} \right| \right).
\end{equation} 
 \noindent Less sensitive to outliers, by this definition $\sigma_{\textrm{\tiny NMAD}}$ is equal to the standard deviation for a Gaussian distribution, directly comparable to other papers which quote $rms$/(1+$z$). 
     
\begin{figure*}[ht!]
  \begin{center}
  \includegraphics[scale=.89, clip, trim=6mm 0mm 0mm 0mm]{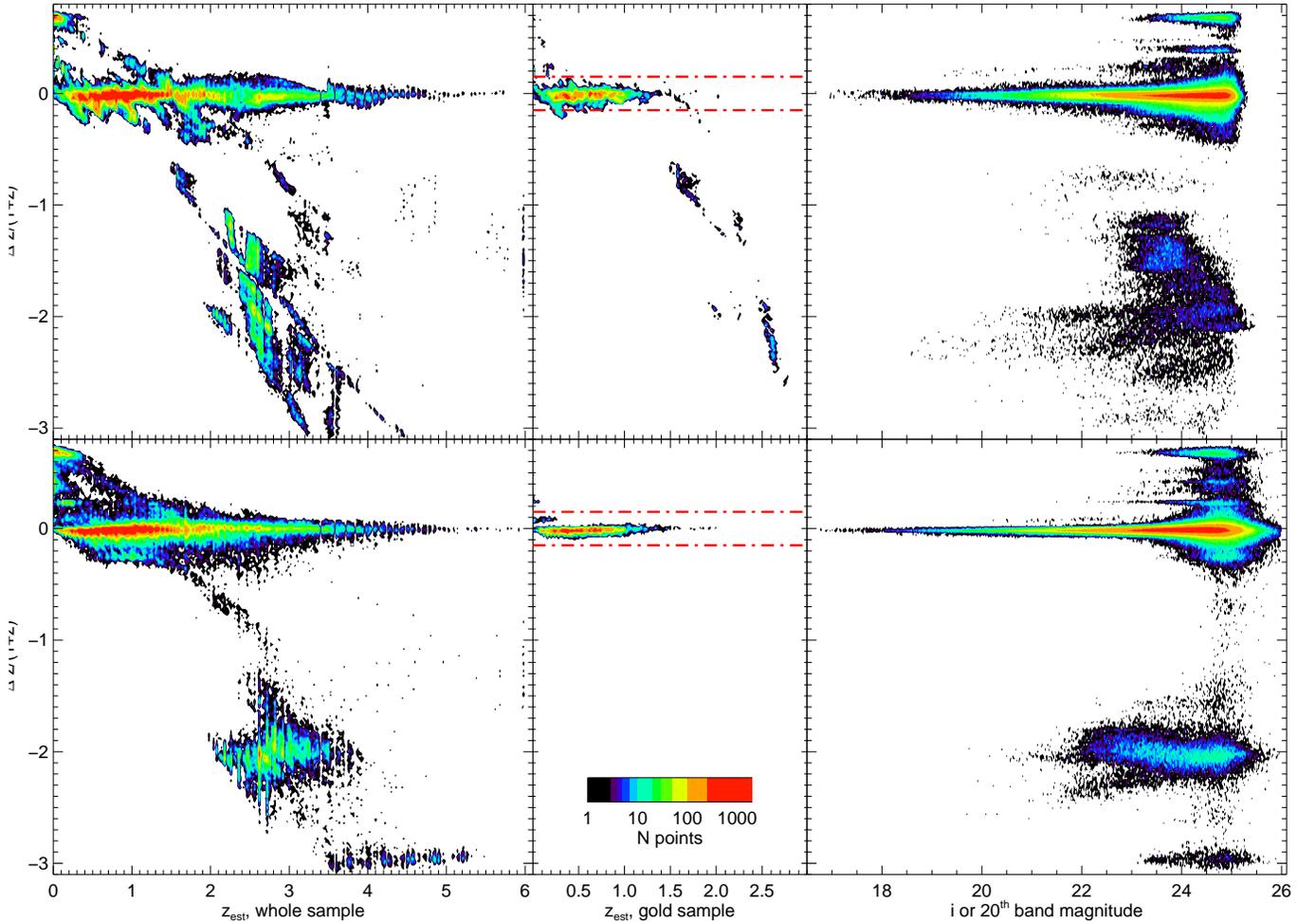} 
     \end{center}
  \caption[.]{Illustration of simulation results for LSST (top row) and Giga-$z$ (bottom row).  The leftmost and rightmost columns show density contours for all objects in the input CMC catalog ($i\,<\,$25), the middle column shows the ``gold sample'' (see Table \ref{tab:zstats}).  A black point indicates a handful of sources in the bin, red points are for $\gtrsim$\,200.  Red dashed lines demarcate catastrophic outliers, where $| \Delta z |/(1+z)$\,$>$\,0.15.  The leftmost column of plots show the scatter in estimated redshift as a function of estimated redshift, and the rightmost column of plots as a function of measured magnitude (in the $i$ band for LSST, or the 20$^{th}$ spectral band for Giga-$z$, which sits at roughly the middle of the $i$ band).  The text of Section \ref{sec:photoz_res} discusses the results in detail.}
  \label{fig:gigazvslsst}
\end{figure*}

Figure \ref{fig:gigazvslsst} shows the distribution of  $\Delta z$/(1+$z_{in}$) for the Pegase13 templates, as a function of the known input catalog redshift and as a function of the object's observed $i$ band magnitude.  The top line of plots are for our LSST simulations, and the bottom line for Giga-$z$.  The horizontal lines at $\pm$\,0.15 demarcate catastrophic failures from the central swath. 

We find that these statistics are most sensitive to the following:

\begin{itemize}\itemsep0pt
\item N$_{filters}$ 
\item Wavelength coverage
\item Redshift
\item S/N 
\item Width of the redshift probability distribution, P($z$)
\item The number of spectral types being fit for
\item Template set
\end{itemize}

{
\renewcommand{\arraystretch}{1.3}

\begin{deluxetable*}{lccccccccc}[h!]
\tablecaption{Redshift recovery statistics for the LSST and Giga-$z$ simulations with various parameter and template set cuts to illustrate their effect. \label{tab:zstats}}
\tablewidth{0pt}

\tablehead{
\colhead{} & \multicolumn{4}{c}{{\bf LSST}} & \colhead{} & \multicolumn{4}{c}{{\bf Giga-$z$}}   \\  
\colhead{\bf Parameter Selection} & 
\colhead{$\sigma_{\rm {\tiny NMAD}}$} & \colhead{Catastrophic\tablenotemark{a}} & \colhead{Bias} & \colhead{\% Catalog} & \colhead{} &
\colhead{$\sigma_{\rm {\tiny NMAD}}$} & \colhead{Catastrophic\tablenotemark{a}} & \colhead{Bias} & \colhead{\% Catalog} \\
\colhead{} &
\colhead{} & \colhead{Failures (\%)} & \colhead{} & \colhead{Remaining} & \colhead{} &
\colhead{} & \colhead{Failures (\%)} & \colhead{} & \colhead{Remaining} 
}                                                          
                                                                       
\startdata
\cutinhead{{\bf Template Set}\tablenotemark{b}} 
EAZY                                                           & 0.061 & 25.4 & -0.011 & 100 & & 0.038 & 22.7 & 0.001 & 100 \\
Pegase13                                                   & 0.041 & 18.4 & -0.014 & 100 & & 0.030 & 18.7 & -0.008 & 100 \\
 \cutinhead{{\bf Magnitude}\tablenotemark{c}} 
$<$\,24.5                                                    & 0.037 & 17.4 & -0.017 & 71.4 & & 0.023 & 15.8 & -0.008 & 64.4 \\
$<$\,24                                                       & 0.035 & 17.2 & -0.016 & 49.8 & & 0.018 & 14.0 & -0.008 & 43.9 \\                                   
$<$\,22.5                                                    & 0.032 & 12.5 & -0.012 & 14.1 & & 0.012 & 10.1 & -0.007 & 12.5 \\                                                   
\cutinhead{{\bf Redshift}} 
0.5\,$<$\,$z$                                               & 0.033 & 10.3 & -0.008 & 76.7 & & 0.026 & 10.3 & -0.002 & 76.7 \\
0.5\,$<$\,$z$\,$<$\,2.25, 3\,$<$\,$z$\,$<$\,6    & 0.031 & 8.7 & -0.008 & 68.7 & & 0.025 & 8.7 & -0.004 & 68.7 \\                                                                             
\cutinhead{{\bf Redshift Probability Distribution Width}} 
$W_{99}$\tablenotemark{d}\, $<$\,2.5       & 0.035 & 15.4 & -0.015 & 77.7 & & 0.023 & 14.1 & -0.007 & 72.2 \\                                                             
\cutinhead{{\bf Combined ``Gold Sample''}}  
mag\,$<$\,22.5, $\overline{P(z)}$\,$>$\,0.17    & 0.029 & 5.4 & -0.008 & 13.0 & & 0.010 & 0.3 & -0.006 & 11.2 \\   
\cutinhead{{\bf Spectral Type}}  
Pegase13 ellipticals subset                      & 0.028 & 2.0 & 0.003 & 2.2 &  & 0.007 & 2.5 & -0.001 & 2.7 \\                                                                           

\enddata 

\tablenotetext{a}{Defined as $| \Delta z |$/(1+$z$)\,$<$\,0.15.}
\tablenotetext{b}{All other quantities shown were calculated using the Pegase13 template set.}
\tablenotetext{c}{The cut on observed magnitude applies for the $i$ band for LSST, or 20$^{th}$ filter band for Giga-$z$, which has a similar central wavelength.}
\tablenotetext{d}{The the 99\%\ confidence width of the posterior redshift distribution from EAZY (u99\,-\,l99, where u99 and l99 are parameters returned by EAZY).}

\end{deluxetable*}
}

{
\renewcommand{\arraystretch}{1.3}

\begin{deluxetable*}{lrrccrc}[h!]
\tablecaption{A comparison of redshift recovery statistics between multi-band photometry or multi-object spectroscopy experiments, both past and planned. \label{tab:compare}}
\tablewidth{0pt}

\tablehead{
\colhead{Experiment} & 
\colhead{N$_{gals}$} & 
\colhead{Area [deg$^2$]} & 
\colhead{Magnitude Limit} &  
\colhead{N$_{filts}$/Resolution} & 
\colhead{Scatter} & 
\colhead{Cat. Failure Rate}
}

\startdata
COMBO 17 \tablenotemark{a}             & $\sim$10,000       & $\sim$0.25   & $R\,<$\,24                            &  17                & 0.06                      & $\lesssim$\,5\% \\   
COSMOS \tablenotemark{b}                &  $\sim$100,000   & 2                      & $i^+_{AB}\,\sim$\,24          & 30                 & 0.06                       & $\sim$\,20\% \\      
                                                                  & $\sim$30,000      & 2                      & $i^+\,<$\,22.5                      & 30                 & 0.007                    &  $<$\,1\%\\  
CFHTLS - Deep \tablenotemark{c}     & 244,701                & 4                     & $i'_{AB}\,<$\,24                   & 5                   & 0.028                    & 3.5\% \\  
CFHTLS - Wide \tablenotemark{c}     & 592,891                & 35                   & $i'_{AB}\,<$\,22.5                & 5                   & 0.036                    & 2.8\% \\                                                 
PRIMUS \tablenotemark{d}                 & 120,000                & 9.1                  & $i_{AB}\,\sim$\,23.5            & R$_{423}$\,$\sim$\,90 & $\sim0.005$    & $\sim$2\%  \\          
WiggleZ \tablenotemark{e}                  & 238,000                & 1,000             & 20\,$<$\,$r$\,$<$\,22.5       & R$_{423}$\,=\,845        & $\lesssim$\,0.001  &  $\lesssim$\,30\%             \\                                                  
Alhambra \tablenotemark{f}                & 500,000                & 4                      & $I\,\leq$\,25                          & 23                & 0.03                       &  \nodata \\
BOSS  \tablenotemark{g}                      & 1,500,000            & 10,000             & $i_{AB}$\,$\leq$\,19.9       & R$_{423}$\,$\sim$\,1600  &  $\lesssim$0.005    & $\sim$2\%  \\ 
DES \tablenotemark{h}                         & 300,000,000       & 5,000                & $r_{AB}\,\lesssim$\,24       &  5                 &   0.1                        &   \nodata \\
EUCLID \tablenotemark{i}                   & 2,000,000,000     & 15,000            &    Y,J,K\,$\lesssim$\,24    & 3$^+$           &   $\lesssim$\,0.05    &       $\lesssim$\,10\%           \\
                                                                  & 50,000,000          & 15,000             &    H$_{\alpha}$\,$\geq$\,3e-16\,erg/s/cm$^2$   & R$_{1\mu m}$\,$\sim$\,250     &       $\lesssim$\,0.001   &     $<$\,20\%             \\          
LSST  \tablenotemark{j}                      & 3,000,000,000    & 20,000             & $i_{AB}\,\lesssim$\,26.5    &  6                &  $\lesssim$\,0.05   & $\lesssim$\,10\% \\                                                                                                                                                                                            
Giga-$z$                                                 & 2,000,000,000    & 20,000             & $i_{AB}\,\lesssim$\,25.0     & R$_{423}$\,=\,30     & 0.03          &  $\sim$\,19\% \\
                                                                  & 224,000,000       & 20,000             & $i_{AB}\,\lesssim$\,22.5     & R$_{423}$\,=\,30     & 0.01          & 0.3\% \\
\enddata
\tablenotetext{a}{ \cite{wolf_2004}}
\tablenotetext{b}{\cite{ilbert_2009}}
\tablenotetext{c}{\cite{coupon_2009}}
\tablenotetext{d}{\cite{coil_2011}; Resolution is per slit width, whereas at 423\,nm, the PRIMUS resolution per pixel is $\approx$\,400.}
\tablenotetext{e}{\cite{drinkwater_2010}; We consider the galaxies observed for an hour without robust redshifts to be failures.}
\tablenotetext{f}{\cite{moles_2008}; Expected.}
\tablenotetext{g}{\cite{dawson_2013} and references therein.}
\tablenotetext{h}{\cite{banerji_2008}; Expected.}
\tablenotetext{i}{\cite{euclid_2011}; Expected.  Photometric redshifts rely on combination of the Y,J,K bands with ground based photometry in 4 visible bands derived from public data or through collaborations.}
\tablenotetext{j}{\cite{lsst_2009}; The quoted number of galaxies that will have photometric redshifts obtained, and LSST quoted scatter and catastrophic failure rate.  See Table \ref{tab:zstats} for the findings from this study.}

\end{deluxetable*}
}

\noindent Therefore we have included in Table \ref{tab:zstats} the effect of making cuts on the parameters not dictated by experimental design.  In particular, we note that as scatter is correlated with how many spectral types one must fit for, increasing the number of objects in the catalog does not reduce the systematic error -- scatter or incidence of catastrophic failures.  However, better sampling statistics will improve the mean uncertainty in each redshift bin.  We tested this by analyzing only a randomly chosen fraction of our catalog, where the fraction was one of [3/4, 2/3, 1/2, 1/3, 1/4].  The statistics presented here are robust to this type of selection to less than 0.1\%.  

As expected, the effect of increasing the MKID detector resolution to R$_{423}$\,=\,60 or increasing the exposure time per target from 15 to 30 minutes decreased both the scatter (by $\approx$\,$\sqrt{2}$) and the catastrophic failure rate.

EAZY allows for the construction of a quality factor, Q$_z$, with each computed redshift that depends on the $\chi^2$ value as well as the 99\% confidence interval and the integrated probability, as a metric for the reliability of the redshift estimates that does not preferentially select out high redshift sources.  However, we found that Q$_z$ was not a good predictor of redshift accuracy.

Many objects had multiply peaked redshift probability distributions, and/or were highly non-Gaussian.  We have made one broad cut on width of the P($z$), but more complicated functions of the probability distribution characteristics such as peak height might be useful to pursue in future studies.  

The estimated redshift results which gave the least scatter were obtained using the Pegase13 templates, fit one at a time because the solution time quickly becomes prohibitively large as the number of templates increases.  However, we found that the overall bias was reduced when using the EAZY templates.  The template-fitting redshift estimation method is susceptible to the fact that template colors and redshift are often degenerate.  Empirical redshift estimation codes may produce smaller biases (by a factor of $\approx$\,2), since the model will match the data better by construction, suggesting systematic inaccuracies in most template sets. A sufficient training set, however, could be used to recalibrate templates, thereby reducing the inaccuracy \citep[e.g.,][]{budavari_2000}.  Spectroscopic calibration samples themselves, however, may lack spectra of some subset of rare galaxies that otherwise may not be easily identified and removed (see e.g. \cite{newman_2008}, for a discussion). 

Contamination occurs predominantly in two ``islands''.   One is in the redshift desert, at 2.25\,$\lesssim$\,$z$\,$\lesssim$\,3 for the Giga-$z$ filters.  The second is at $z_{in}$\,$\lesssim$\,0.7.  These are most likely caused by attributing the high redshift Lyman break to the low redshift 4000\,\AA\ break, and vice versa.  Of course, SEDs not well represented by the template set will have issues, as will very blue galaxies with featureless SEDs.   This simulation does not incorporate a magnitude prior, although doing so may reduce the size of these islands.

Our findings for LSST of scatters of $\approx$\,3--4\% in $\Delta z$/(1+$z$) are consistent with their studies \citep{lsst_2009}, although we find higher outlier rates, ranging from $\approx$\,5--20\% except for the most extreme cuts, more in alignment with the findings of \cite{hildebrandt_2010}.  The redshift estimates for Giga-$z$ are superior, in terms of dispersion, bias, and catastrophic failure rate by up to a factor of over 3 over all parameter cuts, highlighting how Giga-$z$'s $\approx$ log spectral resolution improves on the commonly used optical filters.  The inclusion of nIR data could improve both the LSST and Giga-$z$ results since it is then possible to simultaneously constrain both the Lyman and 4000\,\AA~ breaks.

Though it is difficult to make strict comparisons with other experiments, we list some of the salient statistics for various similar multi-band photometric or multi-object spectroscopic experiments, both past and planned, in Table \ref{tab:compare}.  Though not comprehensive, it gives an idea of the state of the field.  Because all Giga-$z$ ``filters'' observe simultaneously, Giga-$z$ does not suffer from the trade-off between photometric depth (or number density) and higher spectral resolution that the experiments using more, narrower filters or spectroscopy will face.
          
Finally, we mention the cleaning of outliers to yield lower outlier rates.  Depending on the science application such a filtering can be effective, for example, for Dark Energy studies with weak lensing that do not rely on complete galaxy samples.  However, some science applications do rely on redshifts for all objects not allowing for filtering. For those kind of applications the results reported in this Section are particularly informative.


\subsection{Luminous Red Galaxies}          
\label{sec:lrgs}

Baryon Acoustic Oscillations (BAO) are ripples that appear in the spatial pattern of galaxies, exhibiting coherence on the particular co-moving scale of $\approx$\,150\,Mpc separation, determined from cosmic microwave background (CMB) observations \citep{komatsu_2011}.  They appear in the galaxy distribution as a ``bump'' \citep{eisenstein_2007}, or ``wiggles'' in the matter fluctuation power spectrum \citep{cole_2005}, analogous to a low redshift CMB power spectrum.  Since the physical size is known, BAO serve as a ruler with which to measure the geometry of the Universe in both the radial and angular directions.  Furthermore, BAO in these orthogonal directions are subject to different systematics, which can be used as a cross check.  Indeed, BAO may have the lowest level of systematic uncertainty of all current Dark Energy probes \citep{albrecht_2007}.

Reaping the potential of BAO, however, requires redshift estimates more accurate than what was found with our main galaxy population in Section \ref{sec:photoz_res}.  This is similarly true for WL analyses, in order to separate galaxies into redshift slices and correct for intrinsic galaxy alignment contamination.  Traditionally, spectroscopy has been used to ascertain galaxy redshifts from which distances are derived, probing cosmology through an averaged three-dimensional BAO measurement.  

More recently, multi-object spectroscopic experiments such as the WiggleZ survey \citep{blake_2011}, HETDEX \citep{hill_2008}, BOSS \citep{schlegel_2009}, and Big BOSS \citep{schlegel_2011} have been conducted or are planned, with the aim of improving on the current BAO measurements.  However, at present, uncertainties in the Dark Energy constraints set by BAO are limited by data volume.  To fully realize the potential of this method, larger numbers of objects (yet with precise redshift estimation) are needed.  Luckily, for WL or BAO Dark Energy analyses, it is not necessary to have complete galaxy samples.  Therefore low resolution galaxy surveys can optimally select a subset of galaxies so as to meet the redshift accuracy criterion for the sample with minimal impact on the derived cosmological constraints \citep[e.g.,][]{padmanabhan_2007}. 

Luminous red galaxies (LRGs) are a homogenous subset of the main galaxy population, predominantly massive early-type galaxies.  They are strongly biased, mapping the observable galaxy distribution to the underlying mass density distrubition.  Intrinsically luminous, they are excellent tracers of large scale structure, and an observational sample can be selected using photometric colors relatively easily to high redshift  due to the strong 4000\,\AA\ break in their SEDs  \citep[e.g.,][]{eisenstein_2001,collister_2007}.

\cite{benitez2_2009} show that $\sigma_{\Delta z /(1+z)}$\,$\approx$\,0.003 is sufficient precision to measure the BAO in the radial direction, which has more stringent requirements than the angular direction.  Doing better than this merely oversamples the BAO bump, and at this level, redshift space distortions and nonlinear effects can produce comparable errors.  This limit corresponds to $\approx$\,15\,Mpc/h, the intrinsic co-moving width of the bump along the line of sight at $z$\,=\,0.5, caused primarily by Silk damping \citep{silk_1968}.  This is the mean redshift for the PAU LRG survey which will cover the northern skies over an area of 8000\,deg$^2$, sampling cosmological volume V\,$\approx$\,10\,h$^{-3}$\,Gpc$^3$ \citep{benitez2_2009}.  By attempting reconstruction, aided by the need for fewer fitting templates, it is possible to do even better, as was done for BOSS \citep{padmanabhan_2012}.

In Section \ref{sec:photoz_res} we showed that limiting the galaxy catalog to a sample of elliptical galaxies reduces the scatter in redshift error to $\sigma_{\rm {\tiny NMAD}}$\,=\,0.007 with a catastrophic failure rate and average bias of 2.5\% and -0.001 respectively.   In practice, LRG selection is based on color and luminosity selection, but this is not usually difficult for LRGs.  With an energy resolution of R$_{423\,\rm nm}$\,=\,30 and the number density of objects catalogued by LSST, Giga-$z$ should be able to achieve the necessary redshift estimation accuracy for LSST LRGs.  The large survey volume probed will ensure an error not dominated by sampling limits, and shot noise should be comparable or less to that of the PAU experiment.  In addition, since Giga-$z$ would probe the southern hemisphere, it allows for a joint analysis of the data sets.  


\subsection{Quasars}  
\label{sec:qsos}

Quasars are extremely luminous objects, believed to be accreting supermassive black holes.   Type I quasars, due to the high velocities of accretion disk material, are observed to have characteristic broad ($\approx$\,1/20--1/10 FWHM) emission lines in their SEDs \citep{vandenberk_2001}.  They are UV dropout objects, and thus broadband filters will only begin to see the Ly-$\alpha$ break ($\lambda_{rest}$\,$\approx$\,1200\,\AA) at $z$\,$\gtrsim$\,2.2. 

Though their number density is small compared to ordinary galaxies, quasars are more biased tracers of the matter distribution, and their bias increases with redshift.  Visible out to high redshifts, the cosmological volume they can be used to probe is much greater than with galaxies, and since sample variance and shot noise decrease as the square root of the volume, they have the potential to measure large scale structure even better than LRGs around the peak of the matter power spectrum in the range $z$\,$\approx$\,1--3.

Furthermore, quasars can also be used to measure BAO at high redshift where systematic effects such as redshift distortions and nonlinearities have less influence.  \cite{sawangwit_2012} used the Sloan Digital Sky Survey (SDSS), 2dF QSO redshift survey (2QZ), and 2dF-SDSS LRG and QSO (2SLAQ) quasar catalogs to measure BAO features, but these were detected only at low statistical significance.  However, they estimate that a quarter million $z$\,$<$\,2.2 quasars over 3000\,deg$^2$ would yield a $\approx$\,3$\sigma$ detection of the BAO peak.

LSST predicts that they will produce a catalog of roughly 10 million quasars.  Likely these will be identified using photometric selection through color-color and color-magnitude diagrams as was done for SDSS \citep{richards_2009}.  However, above $z$\,$\approx$\,2.5, selection becomes much more difficult as quasar colors become indistinguishable from that of stars.

\begin{figure}[h!]
  \begin{center}
  \includegraphics[scale=0.43, clip, trim=10mm 5mm 0mm 6mm]{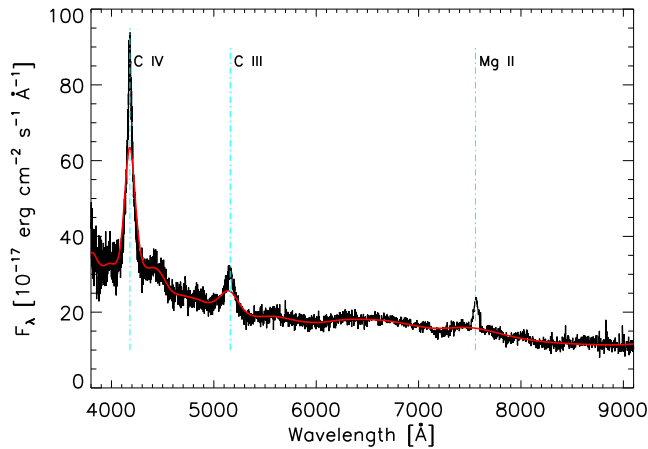}
  \end{center}
  \caption[quasar.]{ SDSS J001507.00-000800.9 quasar spectrum in black, at the resolution of Giga-$z$ in red. This QSO has a redshift of 1.703. }
  \label{fig:quasar}
\end{figure}

With the R$_{423\,\rm nm}$\,=\,30 resolution of Giga-$z$, the broad emission lines of quasars will be resolved, and their redshifts can be estimated using these spectral features.  The high number counts from LSST and redshift accuracy enabled by Giga-$z$ could, with negligible cost, provide low resolution spectroscopy for the LSST quasar candidates, yielding a precision measurement of the matter power spectrum as well as BAO at high redshift.  Furthermore, quasars could be unambiguously told apart from stars.  Besides BAO and measurements of the distribution of structure to $z$\,$\approx$\,6, Giga-$z$ could also be used to study the quasar luminosity function, quasar clustering and bias, set limits on the quasar duty cycle, and improve our understanding of these objects and their evolution and co-evolution with their host galaxies.

While we do not explicitly predict the redshift accuracies achievable with Giga-$z$ here, leaving it for a future investigation, \cite{abramo_2011} show that with the J-PAS instrument, with 42 contiguous 118\,\AA\ FWHM filters spanning 430--815\,nm, they could extract photo-$z$'s of type I quasars with (rms) accuracy $\sigma_{\Delta z}$\,$\approx$\,$0.001(1+z)$.  They show that it is possible to obtain near-spectroscopic photometric redshifts (which suffer from intrinsic errors due to line shifts) for quasars with a template fitting method, with a negligible number of catastrophic redshift errors.  Higher resolution spectra or greater S/N would mostly serve to bring down the number of catastrophic errors.


\section{Forecasted Cosmological Constraints for LSST and Giga-$z$}
\label{sec:de}

We now estimate the impact of the Giga-$z$ photometric redshifts on constraints from weak lensing with LSST. In general such a forecast requires the selection of a galaxy sample and a model for  statistical and systematic errors. We describe both of these, presenting results appropriate for 3 years of data for both instruments.

\subsection{Galaxy Sample Construction}

We first construct the subset of the COSMOS Mock Catalog that has successful shape measurements and photo-$z$s. There are several options for doing this with varying levels of aggressiveness depending on S/N cuts, definition of a ``resolved'' galaxy, and photo-$z$ quality.

To construct the galaxy sample, we first consider the objects resolved by LSST that would have measured shapes. The specific cuts applied were:

\begin{list}{$\bullet$}{\itemsep 2pt}
\item The resolution factor Res\,$>$\,0.4. The resolution factor is defined using the \citet{bernstein_2002} convention:
\begin{equation}
{\rm Res} = \frac{r_{\rm 1/2,gal}^2}{r_{\rm 1/2,gal}^2 + r_{\rm 1/2,psf}^2}\,,
\end{equation}
where $r_{\rm 1/2,gal}$ and $r_{\rm 1/2,psf}$ are the half-light radii of the galaxy and the point-spread function (PSF) respectively.  This cut prevents source galaxies small compared to the PSF from being used.
\item The detection signal-to-noise ratio S/N\,$>$\,18.
\item The ellipticity measurement uncertainty $\sigma_e$\,$<$\,0.2 (per component). Again the ellipticity definition (and calculation of $\sigma_e$) follows \citet{bernstein_2002}: $e$\,=\,$(a^2-b^2)/(a^2+b^2)$, where $a$ and $b$ are the major and minor axes.
\end{list}

\noindent Alternative cuts are possible, which may lead to a larger or smaller source sample.

For LSST, we assumed $r_{\rm 1/2,psf}$\,=\,0.39\,arcsec, which is obtained for a Kolmogorov profile with a full width at half maximum of 0.69\,arcsec. The depth of the imaging data was taken to be $r$\,=\,26.8 and $i$\,=\,26.2 (5$\sigma$ point source; this is after 3 years, the final LSST dataset will be deeper). Galaxy shape catalogs were generated separately in the $r$ and $i$ filters, and their union taken. The  density of objects with successful shape measurements is 14.9\,gal/arcmin$^2$.

Not all of the objects with successful shape measurements can be used: they must also have reliable photo-$z$s. To account for this, we split the galaxies into photo-$z$ bins of  width $\Delta z_{\rm est}$\,=\,0.1. If photo-$z$s were always a good tracer of the true redshift, we could use all of the galaxies in each bin. In practice, the redshift outliers must be removed as they  contribute to pernicious systematic errors, even if the probability distribution $P(z_{\rm in}|z_{\rm est})$ were exactly known. For example, intrinsic galaxy alignments -- which are known to exist for red  galaxies and may contaminate the true lensing signal at the level of up to 2--3\%\ for blue galaxies \citep{mandelbaum_2006, hirata_2007, mandelbaum_2011} -- can in principle be removed via the redshift  dependence of the signal \citep[e.g.][]{takada_2004, hirata_2004, king_2005, kirk_2010}. However, these removal methods do not work if the redshift outliers are themselves intrinsically aligned. The only safe  approach is to reduce the outlier rate.  Thus we impose a requirement on the outlier rate of $<$\,5\%, so that if their intrinsic alignments are $\sim$\,2\,\%\ of the lensing signal, the overall contamination is at no more than the part-per-thousand level. This requirement is achieved by the following method:

\begin{list}{$\bullet$}{\itemsep 2pt}
\item For each galaxy, we compute $W_{99}$, the 99\%\ confidence width of the posterior redshift distribution from EAZY.
\item In each photo-$z$ bin, we impose a cut $W_{99,\rm max}$ on $W_{99}$. This cut is reduced until the 5\%\ outlier rate is met.
\item In some cases, no cut on $W_{99}$ can reduce the outlier rate below 5\%: in this case that photo-$z$ bin is completely removed from the sample.
\end{list}

This results in a culled galaxy catalog with both measured shapes and reliable photo-$z$s. The size of the culled catalog increases as the photo-$z$ performance improves. The unweighted density of  galaxies $n$ in this catalog is 12.2\,gal/arcmin$^2$ (LSST photo-$z$) versus 13.0\,gal/arcmin$^2$ (Giga-$z$ photo-$z$).

Once the galaxy catalog is constructed, the effective source density is obtained via

\begin{equation}
n_{\rm eff} = \sum_j \frac1{1+(\sigma_{e,j}/0.4)^2}\,,
\end{equation}

\noindent where the sum is over galaxies in the catalog, and the factor in the sum down-weights galaxies whose measurement uncertainty is significant compared with the intrinsic RMS dispersion of $\sim$\,0.4. The effective source densities are 14.5 (all shapes), 11.9 (LSST photo-$z$), and 12.7 (Giga-$z$ photo-$z$) gal/arcmin$^2$.

\subsection{Parameter Forecasting Methodology}

Cosmological parameter constraints were estimated using the weak lensing Fisher matrix code from the Figure of Merit Science Working Group \citep[FoMSWG;][]{albrecht_2009}. This code includes constraints  from the shear power spectrum and the geometrical part of galaxy-galaxy lensing (i.e. ratios of the signals at various redshifts, which depend only on the background cosmology and not the relationship  between the galaxies and the mass; \citealt{bernstein_2004}). The inner workings are described at length in the FoMSWG report \citep[][Appendix A2]{albrecht_2009} and will not be repeated here except for the intrinsic alignment models, which have been updated. In addition  to statistical errors, the FoMSWG code enables several systematic errors to be included: (i) shear calibration errors; (ii) photo-$z$ biases; (iii) non-Gaussian contributions to the covariance matrix of  lensing power spectra; and (iv) intrinsic alignments. We turn off (iii) here since recent studies have suggested that the effect can be mitigated by nonlinear transformations on the shear map that remove the non-Gaussian tails of the lensing convergence distribution contributed by the most massive halos \citep[e.g.][]{seo_2011}. 

The original FoMSWG forecasting software did not include photo-$z$ outliers and so we revisit the issue here. FoMSWG assumed that the systematic uncertainty in the photo-$z$s could be captured in a  photo-$z$ bias $\delta z_{{\rm est},i}$ where $i$\,=\,$1...N_z$ indicates a redshift slice index. FoMSWG then assumed that a complete spectroscopic survey of $N_{\rm spec}$ source galaxies would be conducted to  calibrate the photo-$z$ error distribution. If the fraction of source galaxies in the $i^{\rm th}$ redshift slice is $f_i$, then it follows that there would be $N_{\rm spec}f_i$ spectroscopic galaxies in the $i^{\rm th}$ slice. Then the spectroscopic survey would enable us to impose a prior on the photo-$z$ bias of width:

\begin{equation}
\sigma_{\rm pr}(\delta z_{{\rm est},i}) = \frac{\sigma_{z,i}}{\sqrt{N_{\rm spec}f_i}}.
\label{eq:prior-pz}
\end{equation}

Here we extend this approach to include a nonparametric description of redshift outliers \citep[e.g.][]{bernstein_2009}. We suppose that a fraction, $\epsilon_{ij}$, of the galaxies in the $i^{\rm th}$ photo-$z$ bin are actually outliers and lie in the $j^{\rm th}$ true redshift bin. Only the true outliers, as defined by $|\Delta z|$\,$>$\,$0.15(1+z)$, are included here -- the ``core'' of the photo-$z$ error  distribution is modeled using the offset parameters $\delta z_{{\rm est},i}$. By construction, $\epsilon_{ii}$\,=\,0 for all $i$ (a galaxy in the correct bin is not an outlier) but note that in general $\epsilon_{ij}$\,$\neq$\,$\epsilon_{ji}$ (the outlier scattering between redshift slices need not be symmetric). The shear power spectrum $C_\ell^{ij}$ between the $i$ and $j$ photo-$z$ slices then differs from the true power spectrum $C_\ell^{ij}({\rm true})$ according to

\begin{eqnarray}
C_\ell^{ij} & = & C_\ell^{ij}({\rm true})+ \sum_k \epsilon_{ik} [C_\ell^{kj}({\rm true})-C_\ell^{ij}({\rm true})]    \nonumber \\
                   &    &  + \sum_k \epsilon_{jk} [C_\ell^{ik}({\rm true})-C_\ell^{ij}({\rm true})],    
\end{eqnarray}

\noindent to first order in $\epsilon_{ik}$. In analogy to Eq.~(\ref{eq:prior-pz}), we treat the suite of $N_z(N_z-1)$ parameters $\{\epsilon_{ij}\}$ as additional nuisance parameters, and impose a prior of the  form

\begin{equation}
\sigma_{\rm pr}(\epsilon_{ij}) = \sqrt{\frac{\epsilon_{ij}}{N_{\rm spec}f_i}}\,.
\label{eq:prior-outlier}
\end{equation}

We consider here both the intrinsic alignment model used by the FoMSWG and several variations. The key issue is the treatment of the correlation between gravitational lensing and intrinsic alignments (the ``GI'' term), which in the FoMSWG was parameterized by the matter-intrinsic ellipticity cross-power spectrum, $P_{me}(k,z) = b_\kappa r_\kappa P_{mm}(k,z)$. In general, the matter density $m$, galaxy density $g$, and galaxy ellipticity $e$ have a joint symmetric $3\times 3$ power spectrum matrix:

\begin{eqnarray}
{\bf P}(k) &=& \left( \begin{array}{ccc} P_{mm}(k) & P_{gm}(k) & P_{me}(k) \\ P_{gm}(k) & P_{gg}(k) & P_{ge}(k) \\
P_{me}(k) & P_{ge}(k) & P_{ee}(k) \end{array} \right)
\nonumber \\
&=& \left( \begin{array}{ccc} 1 & b_gr_g & b_\kappa r_\kappa \\ b_gr_g & b_g^2 & b_gb_\kappa r_{g\kappa} \\
b_\kappa r_\kappa & b_gb_\kappa r_{g\kappa} & b_\kappa^2 \end{array} \right) P_{mm}(k),
\end{eqnarray}

\noindent where the $b$s represent bias cofficients and $r$s represent correlation coefficients. Then $b_\kappa r_\kappa$ is a function of $k$ and $z$. Alternatively, since $k$ maps into a given multipole in accordance with the Limber formula $k = \ell/D(z)$, they may be considered to take on values in each of the $N_\ell$ angular scale bins and each of the $N_z$ redshift bins. The FoMSWG default model (``Model III'' here) imposed a prior on $b_\kappa r_\kappa$ of the following form:

\begin{equation}
\sigma_{\rm pr}(b_\kappa r_\kappa) = \left\{ \begin{array}{lll} 0 & & \ell<300 \\ 0.003\sqrt{N_{\ell,\rm nonlin}(N_z-1)} & & \ell>300 \end{array} \right..
\label{eq:spr}
\end{equation}

\noindent That is, FoMSWG assumed that at large scales (linear scales, roughly $\ell<300$) galaxies trace the matter density well enough that the observable $P_{ge}(k)$ could be used to estimate $P_{me}(k)$ and remove the GI signal. However at the nonlinear scales, a weak prior was applied to prevent $|b_\kappa r_\kappa|$ from exceeding the observed value of $\sim 0.003$ estimated from SDSS \citep{hirata_2007}; also now see the WiggleZ survey results \citep[Table 4]{mandelbaum_2011}, which combined with SDSS give $b_\kappa r_\kappa = 0.003\pm 0.004$ ($1\sigma$) at $z=0.3$ for blue galaxies.\footnote{This is for the fit presented with the redshift dependence exponent $\eta_{\rm other}=0$; as noted there the passive evolution model would predict $\eta_{\rm other}=-2$. Note that the error bar presented in \citet{mandelbaum_2011} is $2\sigma$.} (The square root of the number of bins is inserted to prevent the prior from being ``averaged down'' over many bins.) The following modifications have been considered here, ordered from optimism to pessimism:
\newcounter{iamodel}
\begin{list}{\Roman{iamodel}:}{\usecounter{iamodel}}
\item The GI term is ignored entirely, i.e. $\sigma_{\rm pr}(b_\kappa r_\kappa)=0$. This is an unrealistically optimistic model, in that it assumes that a combination of theoretical modeling and observations of the galaxy density-galaxy ellipticity correlation will allow us to compute and subtract off the GI term even in the nonlinear regime and at all redshifts.
\item The FoMSWG default model, but with the prior coefficient reduced from 0.003 to 0.001. This model assumes that {\em either} further studied will reduce the upper limits on intrinsic alignments for blue galaxies, {\em or} that advances in modeling galaxy bias in the nonlinear regime will allow us to convert $P_{ge}(k)$ into $P_{me}(k)$ with $\sim 30$\%\ uncertainty.
\item The FoMSWG default model.
\item The FoMSWG default model, but with the prior coefficient increased to 10 to effectively force future WL data to constrain the GI power spectrum and its redshift evolution in the nonlinear regime, while simultaneously fitting the cosmological parameters.
\item No prior on $b_\kappa r_\kappa$ is applied: the future WL data must now constrain the full GI power spectrum at all scales and redshifts, while simultaneously fitting the cosmological parameters. This model is designed to be overly pessimistic.
\end{list}
In light of the rapidly improving observational constraints on intrinsic alignments, and the substantial modeling effort in both intrinsic alignments and galaxy biasing, Model II may be a reasonable (though not assured) forecast for the theoretical uncertainty by the time of the LSST weak lensing analysis.

\subsection{Results}
\label{subsec:deres}

  Table \ref{tab:deresults} shows the results for the LSST and Giga-$z$ weak lensing cases including intrinsic alignment terms, assuming a training sample of 25,000 spectroscopic redshifts, and galaxy shapes from LSST photometry.  $\sigma(w_p)$ is the uncertainty on $w$ if it is assumed to be a constant, $\sigma(w_a)$ the uncertainty on the rate of change of $w$.  $\Delta_{\gamma}$ parameterizes the rate of the growth of structure, and $\Omega_k$ the curvature of spacetime. The Planck CMB data are included in all of the models shown, since without the CMB constraints the ``standard'' cosmological parameters ($\Omega_m\,h^2$, $n_s$, ...) can be altered to accommodate a fit of the WL data with almost any smooth dark energy model.

{
\renewcommand{\arraystretch}{1.3}

\begin{table}[ht!]
\begin{center}
\caption{The WL+CMB dark energy parameter constraints.\\The primary CMB is from Planck, the weak lensing shapes from LSST, and the photometric redshifts are from either LSST or Giga-z. Results are shown for all 5 intrinsic alignment models described in the text, ranging from I (very optimistic) to V (very pessimistic). \label{tab:deresults}}
\begin{tabular}{lcccc}
\tableline
\tableline
			    &  $\sigma(w_p)$  \B & $\sigma(w_a)$  \B  &  $\sigma(\Delta_{\gamma})$  \B  &  $\sigma(\Omega_k)$  \B       \\
\tableline
                                & \multicolumn{4}{c}{{\bf IA Model I}}      \\      
\tableline
LSST photo-z         & \B 0.0271 & 0.494 & 0.158 & 0.0246  \\
Giga-z photo-z          & \B 0.0246 & 0.405 & 0.124 & 0.0200 \\
\tableline
                                & \multicolumn{4}{c}{{\bf IA Model II}}      \\      
\tableline
LSST photo-z         & \B 0.0373 & 0.671 & 0.204 & 0.0251  \\
Giga-z photo-z          & \B 0.0341 & 0.562 & 0.157 & 0.0204 \\
\tableline
                                & \multicolumn{4}{c}{{\bf IA Model III}}      \\      
\tableline
LSST photo-z         & \B 0.0382 & 0.695 & 0.221 & 0.0252  \\
Giga-z photo-z          & \B 0.0348 & 0.576 & 0.168 & 0.0205 \\
\tableline
                                & \multicolumn{4}{c}{{\bf IA Model IV}}      \\      
\tableline                                          
LSST photo-z         & \B 0.0396 & 0.743 & 0.273 & 0.0258  \\
Giga-z photo-z          & \B 0.0364 & 0.627 & 0.206 & 0.0211 \\
\tableline
                                & \multicolumn{4}{c}{{\bf IA Model V}}      \\      
\tableline
LSST photo-z         & \B 0.0503 & 1.053 & 0.330 & 0.0270  \\
Giga-z photo-z          & \B 0.0450 & 0.912 & 0.279 & 0.0223 \\
\tableline
\end{tabular}
\end{center}
\end{table}
}

For the ``FoMSWG default'' model (Model III), the improvement in parameter constraints from the Giga-$z$ photo-$z$s for WL+Planck is equivalent to multiplying the LSST Fisher matrix by a factor of $\alpha=1.27$ ($w_p$), 1.53 ($w_a$), or 1.98 ($\Delta_{\gamma}$). Note that e.g. for $w_a$ the improvement is equivalent to both multiplying the LSST coverage area and the training sets by 1.53 and reducing all systematics by a factor of $1/\sqrt{1.53}$. In general, we see larger improvements for the cases with more complex redshift dependence, such as changing $w$ or the rate of growth of structure. The improvement for the rate of growth of structure -- one of the most important constraints for weak lensing since it cannot be probed by supernovae -- is particularly impressive: in Model III, adding the Giga-$z$ photometric redshifts would be as valuable for $\Delta_\gamma$ as adding a Northern Hemisphere LSST.

While the choice of intrinsic alignment model affects the cosmological constraints (as expected, the more optimistic models that assume better knowledge of the intrinsic alignments lead to smaller error bars), the advantages of Giga-$z$ are robust to even the more extreme models presented here. For $\Delta_\gamma$, the ``advantage factor'' $\alpha$ defined in the previous paragraph varies as 1.82, 2.00, 1.98, 1.94, or 1.49 (Models I--V respectively); for $w_a$ it is 1.60, 1.51, 1.53, 1.45, or 1.38 (Models I--V respectively).


\section{Conclusion}     
\label{sec:conc}

Several Dark Energy probes rely on photometric redshift estimates that are accurate and exhibit little bias.  The DES, LSST, EUCLID, KIDS and other wide field imaging experiments will survey much of the sky in the usual photometric bands, but to fully realize their potential, the photo-$z$ scatter and biases must be well calibrated.  We have simulated realistic observations with both a 3 year LSST stack and a proposed experiment, Giga-$z$, and compared the results side-by-side.  The mock catalog used, based on COSMOS observations, is deep, complete, and representative of the real span of galaxies we might expect these experiments to observe, including objects from the ``redshift desert''.  By construction, this mock catalog likely contains objects with unusual SEDs.

We have shown that Giga-$z$, with R$_{423\,\rm nm}$\,=\,30 spectral resolution, spatial resolution, and continuous wavelength coverage between 350 and 1350\,nm can efficiently and effectively obtain spectrophotometry of a much larger and deeper sample of galaxies than is possible with current spectrographs. From our simulations, we predict redshift estimate accuracies of $\sigma_{\Delta z /(1+z)}$\,$\approx$\,0.03 for the whole sample, and $\sigma_{\Delta z /(1+z)}$\,$\approx$\,0.007 for a select subset, which in turn adds constraint on Dark Energy parameters for WL + CMB (Planck).  In particular, for the rate of growth of structure, one of the most important constraints for weak lensing since it cannot be probed by supernovae, for the default FoMSWG model, adding the Giga-$z$ photometric redshifts would be equivalent to doubling the LSST footprint (e.g., by running a second complete LSST survey in the North).  This data could be obtained inexpensively compared with most current and future surveys.   With DES set to come online imminently, Giga-$z$ would be able to use DES catalogs to inform a first pass, and operate in parallel with LSST and other wide field imaging surveys.


\section*{Acknowledgments}    

D.~M. was supported by a grant from the Keck Institute for Space Studies.  C.~H. was supported by DOE DOE.DE-SC0006624, and the David and Lucile Packard Foundation.

The authors would like to thank G. Brammer, J. Zoubian, T. Treu, and W. Spinella for their assistance.



\begin{thebibliography}{89}
\expandafter\ifx\csname natexlab\endcsname\relax\def\natexlab#1{#1}\fi

\bibitem[{{Abramo} {et~al.}(2012){Abramo}, {Strauss}, {Lima},
  {Hern{\'a}ndez-Monteagudo}, {Lazkoz}, {Moles}, {de Oliveira}, {Sendra},
  {Sodr{\'e}}, \& {Storchi-Bergmann}}]{abramo_2011}
{Abramo}, L.~R., {et~al.} 2012, \mnras, 423, 3251

\bibitem[{{Albrecht} \& {Bernstein}(2007)}]{albrecht_2007}
{Albrecht}, A., \& {Bernstein}, G. 2007, \prd, 75, 103003

\bibitem[{{Albrecht} {et~al.}(2009){Albrecht}, {Amendola}, {Bernstein},
  {Clowe}, {Eisenstein}, {Guzzo}, {Hirata}, {Huterer}, {Kirshner}, {Kolb}, \&
  {Nichol}}]{albrecht_2009}
{Albrecht}, A., {et~al.} 2009, ArXiv e-prints

\bibitem[{{Amiaux} {et~al.}(2012){Amiaux}, {Scaramella}, {Mellier}, {Altieri},
  {Burigana}, {Da Silva}, {Gomez}, {Hoar}, {Laureijs}, {Maiorano},
  {Magalh{\~a}es Oliveira}, {Renk}, {Saavedra Criado}, {Tereno},
  {Augu{\`e}res}, {Brinchmann}, {Cropper}, {Duvet}, {Ealet}, {Franzetti},
  {Garilli}, {Gondoin}, {Guzzo}, {Hoekstra}, {Holmes}, {Jahnke}, {Kitching},
  {Meneghetti}, {Percival}, \& {Warren}}]{euclid_2011}
{Amiaux}, J., {et~al.} 2012, in Society of Photo-Optical Instrumentation
  Engineers (SPIE) Conference Series, Vol. 8442, Society of Photo-Optical
  Instrumentation Engineers (SPIE) Conference Series

\bibitem[{{Arnouts} {et~al.}(2002){Arnouts}, {Moscardini}, {Vanzella},
  {Colombi}, {Cristiani}, {Fontana}, {Giallongo}, {Matarrese}, \&
  {Saracco}}]{arnouts_2002}
{Arnouts}, S., {et~al.} 2002, \mnras, 329, 355

\bibitem[{{Bacon} {et~al.}(2005){Bacon}, {Taylor}, {Brown}, {Gray}, {Wolf},
  {Meisenheimer}, {Dye}, {Wisotzki}, {Borch}, \& {Kleinheinrich}}]{bacon_2005}
{Bacon}, D.~J., {et~al.} 2005, \mnras, 363, 723

\bibitem[{{Banerji} {et~al.}(2008){Banerji}, {Abdalla}, {Lahav}, \&
  {Lin}}]{banerji_2008}
{Banerji}, M., {Abdalla}, F.~B., {Lahav}, O., \& {Lin}, H. 2008, \mnras, 386,
  1219

\bibitem[{{Ben{\'{\i}}tez}(2000)}]{benitez_2000}
{Ben{\'{\i}}tez}, N. 2000, \apj, 536, 571

\bibitem[{{Ben{\'{\i}}tez} {et~al.}(2009{\natexlab{a}}){Ben{\'{\i}}tez},
  {Gazta{\~n}aga}, {Miquel}, {Castander}, {Moles}, {Crocce},
  {Fern{\'a}ndez-Soto}, {Fosalba}, {Ballesteros}, {Campa}, {Cardiel-Sas},
  {Castilla}, {Crist{\'o}bal-Hornillos}, {Delfino}, {Fern{\'a}ndez},
  {Fern{\'a}ndez-Sopuerta}, {Garc{\'{\i}}a-Bellido}, {Lobo}, {Mart{\'{\i}}nez},
  {Ortiz}, {Pacheco}, {Paredes}, {Pons-Border{\'{\i}}a}, {S{\'a}nchez},
  {S{\'a}nchez}, {Varela}, \& {de Vicente}}]{benitez2_2009}
{Ben{\'{\i}}tez}, N., {et~al.} 2009{\natexlab{a}}, \apj, 691, 241

\bibitem[{{Ben{\'{\i}}tez} {et~al.}(2009{\natexlab{b}}){Ben{\'{\i}}tez},
  {Moles}, {Aguerri}, {Alfaro}, {Broadhurst}, {Cabrera-Ca{\~n}o}, {Castander},
  {Cepa}, {Cervi{\~n}o}, {Crist{\'o}bal-Hornillos}, {Fern{\'a}ndez-Soto},
  {Gonz{\'a}lez Delgado}, {Infante}, {M{\'a}rquez}, {Mart{\'{\i}}nez},
  {Masegosa}, {Del Olmo}, {Perea}, {Prada}, {Quintana}, \&
  {S{\'a}nchez}}]{benitez_2009}
---. 2009{\natexlab{b}}, \apjl, 692, L5

\bibitem[{{Bernstein} \& {Huterer}(2010)}]{bernstein_2010}
{Bernstein}, G., \& {Huterer}, D. 2010, \mnras, 401, 1399

\bibitem[{{Bernstein} \& {Jain}(2004)}]{bernstein_2004}
{Bernstein}, G., \& {Jain}, B. 2004, \apj, 600, 17

\bibitem[{{Bernstein}(2009)}]{bernstein_2009}
{Bernstein}, G.~M. 2009, \apj, 695, 652

\bibitem[{{Bernstein} \& {Jarvis}(2002)}]{bernstein_2002}
{Bernstein}, G.~M., \& {Jarvis}, M. 2002, \aj, 123, 583

\bibitem[{{Blaizot} {et~al.}(2005){Blaizot}, {Wadadekar}, {Guiderdoni},
  {Colombi}, {Bertin}, {Bouchet}, {Devriendt}, \& {Hatton}}]{blaizot_2005}
{Blaizot}, J., {Wadadekar}, Y., {Guiderdoni}, B., {Colombi}, S.~T., {Bertin},
  E., {Bouchet}, F.~R., {Devriendt}, J.~E.~G., \& {Hatton}, S. 2005, \mnras,
  360, 159

\bibitem[{{Blake} {et~al.}(2011){Blake}, {Kazin}, {Beutler}, {Davis},
  {Parkinson}, {Brough}, {Colless}, {Contreras}, {Couch}, {Croom}, {Croton},
  {Drinkwater}, {Forster}, {Gilbank}, {Gladders}, {Glazebrook}, {Jelliffe},
  {Jurek}, {Li}, {Madore}, {Martin}, {Pimbblet}, {Poole}, {Pracy}, {Sharp},
  {Wisnioski}, {Woods}, {Wyder}, \& {Yee}}]{blake_2011}
{Blake}, C., {et~al.} 2011, \mnras, 418, 1707

\bibitem[{{Blanton} \& {Roweis}(2007)}]{blanton_2007}
{Blanton}, M.~R., \& {Roweis}, S. 2007, \aj, 133, 734

\bibitem[{{Brammer} {et~al.}(2008){Brammer}, {van Dokkum}, \&
  {Coppi}}]{brammer_2008}
{Brammer}, G.~B., {van Dokkum}, P.~G., \& {Coppi}, P. 2008, \apj, 686, 1503

\bibitem[{{Budav{\'a}ri}(2009)}]{budavari_2009}
{Budav{\'a}ri}, T. 2009, \apj, 695, 747

\bibitem[{{Budav{\'a}ri} {et~al.}(2000){Budav{\'a}ri}, {Szalay}, {Connolly},
  {Csabai}, \& {Dickinson}}]{budavari_2000}
{Budav{\'a}ri}, T., {Szalay}, A.~S., {Connolly}, A.~J., {Csabai}, I., \&
  {Dickinson}, M. 2000, \aj, 120, 1588

\bibitem[{{Calzetti}(2001)}]{calzetti_2001}
{Calzetti}, D. 2001, New Astronomy Reviews, 45, 601

\bibitem[{{Capak} {et~al.}(2007){Capak}, {Aussel}, {Ajiki}, {McCracken},
  {Mobasher}, {Scoville}, {Shopbell}, {Taniguchi}, {Thompson}, {Tribiano},
  {Sasaki}, {Blain}, {Brusa}, {Carilli}, {Comastri}, {Carollo}, {Cassata},
  {Colbert}, {Ellis}, {Elvis}, {Giavalisco}, {Green}, {Guzzo}, {Hasinger},
  {Ilbert}, {Impey}, {Jahnke}, {Kartaltepe}, {Kneib}, {Koda}, {Koekemoer},
  {Komiyama}, {Leauthaud}, {Le Fevre}, {Lilly}, {Liu}, {Massey}, {Miyazaki},
  {Murayama}, {Nagao}, {Peacock}, {Pickles}, {Porciani}, {Renzini}, {Rhodes},
  {Rich}, {Salvato}, {Sanders}, {Scarlata}, {Schiminovich}, {Schinnerer},
  {Scodeggio}, {Sheth}, {Shioya}, {Tasca}, {Taylor}, {Yan}, \&
  {Zamorani}}]{capak_2007}
{Capak}, P., {et~al.} 2007, \apjs, 172, 99

\bibitem[{{Coe} {et~al.}(2006){Coe}, {Ben{\'{\i}}tez}, {S{\'a}nchez}, {Jee},
  {Bouwens}, \& {Ford}}]{coe_2006}
{Coe}, D., {Ben{\'{\i}}tez}, N., {S{\'a}nchez}, S.~F., {Jee}, M., {Bouwens},
  R., \& {Ford}, H. 2006, \aj, 132, 926

\bibitem[{{Coil} {et~al.}(2011){Coil}, {Blanton}, {Burles}, {Cool},
  {Eisenstein}, {Moustakas}, {Wong}, {Zhu}, {Aird}, {Bernstein}, {Bolton}, \&
  {Hogg}}]{coil_2011}
{Coil}, A.~L., {et~al.} 2011, \apj, 741, 8

\bibitem[{{Cole} {et~al.}(2005){Cole}, {Percival}, {Peacock}, {Norberg},
  {Baugh}, {Frenk}, {Baldry}, {Bland-Hawthorn}, {Bridges}, {Cannon}, {Colless},
  {Collins}, {Couch}, {Cross}, {Dalton}, {Eke}, {De Propris}, {Driver},
  {Efstathiou}, {Ellis}, {Glazebrook}, {Jackson}, {Jenkins}, {Lahav}, {Lewis},
  {Lumsden}, {Maddox}, {Madgwick}, {Peterson}, {Sutherland}, \&
  {Taylor}}]{cole_2005}
{Cole}, S., {et~al.} 2005, \mnras, 362, 505

\bibitem[{{Collister} {et~al.}(2007){Collister}, {Lahav}, {Blake}, {Cannon},
  {Croom}, {Drinkwater}, {Edge}, {Eisenstein}, {Loveday}, {Nichol}, {Pimbblet},
  {de Propris}, {Roseboom}, {Ross}, {Schneider}, {Shanks}, \&
  {Wake}}]{collister_2007}
{Collister}, A., {et~al.} 2007, \mnras, 375, 68

\bibitem[{{Conti} {et~al.}(2001){Conti}, {Mattaini}, {Chiappetti}, {Maccagni},
  {Sant'Ambrogio}, {Bottini}, {Garilli}, {Le F{\`e}vre}, {Saisse}, {Vo{\"e}t},
  {Caputi}, {Cascone}, {Mancini}, {Mancini}, {Perrotta}, {Schipani}, \&
  {Vettolani}}]{conti_2001}
{Conti}, G., {et~al.} 2001, \pasp, 113, 452

\bibitem[{{Coupon} {et~al.}(2009){Coupon}, {Ilbert}, {Kilbinger}, {McCracken},
  {Mellier}, {Arnouts}, {Bertin}, {Hudelot}, {Schultheis}, {Le F{\`e}vre}, {Le
  Brun}, {Guzzo}, {Bardelli}, {Zucca}, {Bolzonella}, {Garilli}, {Zamorani},
  {Zanichelli}, {Tresse}, \& {Aussel}}]{coupon_2009}
{Coupon}, J., {et~al.} 2009, \aap, 500, 981

\bibitem[{{Dawson} {et~al.}(2013){Dawson}, {Schlegel}, {Ahn}, {Anderson},
  {Aubourg}, {Bailey}, {Barkhouser}, {Bautista}, {Beifiori}, {Berlind},
  {Bhardwaj}, {Bizyaev}, {Blake}, {Blanton}, {Blomqvist}, {Bolton}, {Borde},
  {Bovy}, {Brandt}, {Brewington}, {Brinkmann}, {Brown}, {Brownstein}, {Bundy},
  {Busca}, {Carithers}, {Carnero}, {Carr}, {Chen}, {Comparat}, {Connolly},
  {Cope}, {Croft}, {Cuesta}, {da Costa}, {Davenport}, {Delubac}, {de Putter},
  {Dhital}, {Ealet}, {Ebelke}, {Eisenstein}, {Escoffier}, {Fan}, {Filiz Ak},
  {Finley}, {Font-Ribera}, {G{\'e}nova-Santos}, {Gunn}, {Guo}, {Haggard},
  {Hall}, {Hamilton}, {Harris}, {Harris}, {Ho}, {Hogg}, {Holder}, {Honscheid},
  {Huehnerhoff}, {Jordan}, {Jordan}, {Kauffmann}, {Kazin}, {Kirkby}, {Klaene},
  {Kneib}, {Le Goff}, {Lee}, {Long}, {Loomis}, {Lundgren}, {Lupton}, {Maia},
  {Makler}, {Malanushenko}, {Malanushenko}, {Mandelbaum}, {Manera}, {Maraston},
  {Margala}, {Masters}, {McBride}, {McDonald}, {McGreer}, {McMahon}, {Mena},
  {Miralda-Escud{\'e}}, {Montero-Dorta}, {Montesano}, {Muna}, {Myers},
  {Naugle}, {Nichol}, {Noterdaeme}, {Nuza}, {Olmstead}, {Oravetz}, {Oravetz},
  {Owen}, {Padmanabhan}, {Palanque-Delabrouille}, {Pan}, {Parejko},
  {P{\^a}ris}, {Percival}, {P{\'e}rez-Fournon}, {P{\'e}rez-R{\`a}fols},
  {Petitjean}, {Pfaffenberger}, {Pforr}, {Pieri}, {Prada}, {Price-Whelan},
  {Raddick}, {Rebolo}, {Rich}, {Richards}, {Rockosi}, {Roe}, {Ross}, {Ross},
  {Rossi}, {Rubi{\~n}o-Martin}, {Samushia}, {S{\'a}nchez}, {Sayres}, {Schmidt},
  {Schneider}, {Sc{\'o}ccola}, {Seo}, {Shelden}, {Sheldon}, {Shen}, {Shu},
  {Slosar}, {Smee}, {Snedden}, {Stauffer}, {Steele}, {Strauss}, {Streblyanska},
  {Suzuki}, {Swanson}, {Tal}, {Tanaka}, {Thomas}, {Tinker}, {Tojeiro},
  {Tremonti}, {Vargas Maga{\~n}a}, {Verde}, {Viel}, {Wake}, {Watson}, {Weaver},
  {Weinberg}, {Weiner}, {West}, {White}, {Wood-Vasey}, {Yeche}, {Zehavi},
  {Zhao}, \& {Zheng}}]{dawson_2013}
{Dawson}, K.~S., {et~al.} 2013, \aj, 145, 10

\bibitem[{{Day} {et~al.}(2003){Day}, {LeDuc}, {Mazin}, {Vayonakis}, \&
  {Zmuidzinas}}]{day_2003}
{Day}, P.~K., {LeDuc}, H.~G., {Mazin}, B.~A., {Vayonakis}, A., \& {Zmuidzinas},
  J. 2003, \nat, 425, 817

\bibitem[{{de Jong} {et~al.}(2012){de Jong}, {Verdoes Kleijn}, {Kuijken}, \&
  {Valentijn}}]{dejong_2012}
{de Jong}, J.~T.~A., {Verdoes Kleijn}, G.~A., {Kuijken}, K.~H., \& {Valentijn},
  E.~A. 2012, Experimental Astronomy, 34

\bibitem[{{De Lucia} \& {Blaizot}(2007)}]{delucia_2007}
{De Lucia}, G., \& {Blaizot}, J. 2007, \mnras, 375, 2

\bibitem[{{Drinkwater} {et~al.}(2010){Drinkwater}, {Jurek}, {Blake}, {Woods},
  {Pimbblet}, {Glazebrook}, {Sharp}, {Pracy}, {Brough}, {Colless}, {Couch},
  {Croom}, {Davis}, {Forbes}, {Forster}, {Gilbank}, {Gladders}, {Jelliffe},
  {Jones}, {Li}, {Madore}, {Martin}, {Poole}, {Small}, {Wisnioski}, {Wyder}, \&
  {Yee}}]{drinkwater_2010}
{Drinkwater}, M.~J., {et~al.} 2010, \mnras, 401, 1429

\bibitem[{{Eisenstein} {et~al.}(2007){Eisenstein}, {Seo}, {Sirko}, \&
  {Spergel}}]{eisenstein_2007}
{Eisenstein}, D.~J., {Seo}, H.-J., {Sirko}, E., \& {Spergel}, D.~N. 2007, \apj,
  664, 675

\bibitem[{{Eisenstein} {et~al.}(2001){Eisenstein}, {Annis}, {Gunn}, {Szalay},
  {Connolly}, {Nichol}, {Bahcall}, {Bernardi}, {Burles}, {Castander},
  {Fukugita}, {Hogg}, {Ivezi{\'c}}, {Knapp}, {Lupton}, {Narayanan}, {Postman},
  {Reichart}, {Richmond}, {Schneider}, {Schlegel}, {Strauss}, {SubbaRao},
  {Tucker}, {Vanden Berk}, {Vogeley}, {Weinberg}, \& {Yanny}}]{eisenstein_2001}
{Eisenstein}, D.~J., {et~al.} 2001, \aj, 122, 2267

\bibitem[{{Fioc} \& {Rocca-Volmerange}(1997)}]{fioc_1997}
{Fioc}, M., \& {Rocca-Volmerange}, B. 1997, \aap, 326, 950

\bibitem[{{Giavalisco} {et~al.}(2004){Giavalisco}, {Ferguson}, {Koekemoer},
  {Dickinson}, {Alexander}, {Bauer}, {Bergeron}, {Biagetti}, {Brandt},
  {Casertano}, {Cesarsky}, {Chatzichristou}, {Conselice}, {Cristiani}, {Da
  Costa}, {Dahlen}, {de Mello}, {Eisenhardt}, {Erben}, {Fall}, {Fassnacht},
  {Fosbury}, {Fruchter}, {Gardner}, {Grogin}, {Hook}, {Hornschemeier}, {Idzi},
  {Jogee}, {Kretchmer}, {Laidler}, {Lee}, {Livio}, {Lucas}, {Madau},
  {Mobasher}, {Moustakas}, {Nonino}, {Padovani}, {Papovich}, {Park},
  {Ravindranath}, {Renzini}, {Richardson}, {Riess}, {Rosati}, {Schirmer},
  {Schreier}, {Somerville}, {Spinrad}, {Stern}, {Stiavelli}, {Strolger},
  {Urry}, {Vandame}, {Williams}, \& {Wolf}}]{giavalisco_2004}
{Giavalisco}, M., {et~al.} 2004, \apjl, 600, L93

\bibitem[{{Grazian} {et~al.}(2006){Grazian}, {Fontana}, {de Santis}, {Nonino},
  {Salimbeni}, {Giallongo}, {Cristiani}, {Gallozzi}, \&
  {Vanzella}}]{grazian_2006}
{Grazian}, A., {et~al.} 2006, \aap, 449, 951

\bibitem[{{Hanuschik}(2003)}]{hanuschik_2003}
{Hanuschik}, R.~W. 2003, \aap, 407, 1157

\bibitem[{{Hearin} {et~al.}(2010){Hearin}, {Zentner}, {Ma}, \&
  {Huterer}}]{hearin_2010}
{Hearin}, A.~P., {Zentner}, A.~R., {Ma}, Z., \& {Huterer}, D. 2010, \apj, 720,
  1351

\bibitem[{{Hildebrandt} {et~al.}(2010){Hildebrandt}, {Arnouts}, {Capak},
  {Moustakas}, {Wolf}, {Abdalla}, {Assef}, {Banerji}, {Ben{\'{\i}}tez},
  {Brammer}, {Budav{\'a}ri}, {Carliles}, {Coe}, {Dahlen}, {Feldmann}, {Gerdes},
  {Gillis}, {Ilbert}, {Kotulla}, {Lahav}, {Li}, {Miralles}, {Purger},
  {Schmidt}, \& {Singal}}]{hildebrandt_2010}
{Hildebrandt}, H., {et~al.} 2010, \aap, 523, A31

\bibitem[{{Hill} {et~al.}(2008){Hill}, {Gebhardt}, {Komatsu}, {Drory},
  {MacQueen}, {Adams}, {Blanc}, {Koehler}, {Rafal}, {Roth}, {Kelz}, {Gronwall},
  {Ciardullo}, \& {Schneider}}]{hill_2008}
{Hill}, G.~J., {et~al.} 2008, in Astronomical Society of the Pacific Conference
  Series, Vol. 399, Panoramic Views of Galaxy Formation and Evolution, ed.
  T.~{Kodama}, T.~{Yamada}, \& K.~{Aoki}, 115

\bibitem[{{Hirata} {et~al.}(2007){Hirata}, {Mandelbaum}, {Ishak}, {Seljak},
  {Nichol}, {Pimbblet}, {Ross}, \& {Wake}}]{hirata_2007}
{Hirata}, C.~M., {Mandelbaum}, R., {Ishak}, M., {Seljak}, U., {Nichol}, R.,
  {Pimbblet}, K.~A., {Ross}, N.~P., \& {Wake}, D. 2007, \mnras, 381, 1197

\bibitem[{{Hirata} \& {Seljak}(2004)}]{hirata_2004}
{Hirata}, C.~M., \& {Seljak}, U. 2004, \prd, 70, 063526

\bibitem[{{Hoaglin} {et~al.}(1983){Hoaglin}, {Mosteller}, \&
  {Tukey}}]{hoaglin_1983}
{Hoaglin}, D.~C., {Mosteller}, F., \& {Tukey}, J.~W. 1983, {Understanding
  robust and exploratory data anlysis}, ed. {Hoaglin, D.~C., Mosteller, F., \&
  Tukey, J.~W.}

\bibitem[{{Huterer} {et~al.}(2006){Huterer}, {Takada}, {Bernstein}, \&
  {Jain}}]{huterer_2006}
{Huterer}, D., {Takada}, M., {Bernstein}, G., \& {Jain}, B. 2006, \mnras, 366,
  101

\bibitem[{{Ilbert} {et~al.}(2006){Ilbert}, {Arnouts}, {McCracken},
  {Bolzonella}, {Bertin}, {Le F{\`e}vre}, {Mellier}, {Zamorani}, {Pell{\`o}},
  {Iovino}, {Tresse}, {Le Brun}, {Bottini}, {Garilli}, {Maccagni}, {Picat},
  {Scaramella}, {Scodeggio}, {Vettolani}, {Zanichelli}, {Adami}, {Bardelli},
  {Cappi}, {Charlot}, {Ciliegi}, {Contini}, {Cucciati}, {Foucaud}, {Franzetti},
  {Gavignaud}, {Guzzo}, {Marano}, {Marinoni}, {Mazure}, {Meneux}, {Merighi},
  {Paltani}, {Pollo}, {Pozzetti}, {Radovich}, {Zucca}, {Bondi}, {Bongiorno},
  {Busarello}, {de La Torre}, {Gregorini}, {Lamareille}, {Mathez}, {Merluzzi},
  {Ripepi}, {Rizzo}, \& {Vergani}}]{ilbert_2006}
{Ilbert}, O., {et~al.} 2006, \aap, 457, 841

\bibitem[{{Ilbert} {et~al.}(2009){Ilbert}, {Capak}, {Salvato}, {Aussel},
  {McCracken}, {Sanders}, {Scoville}, {Kartaltepe}, {Arnouts}, {Le Floc'h},
  {Mobasher}, {Taniguchi}, {Lamareille}, {Leauthaud}, {Sasaki}, {Thompson},
  {Zamojski}, {Zamorani}, {Bardelli}, {Bolzonella}, {Bongiorno}, {Brusa},
  {Caputi}, {Carollo}, {Contini}, {Cook}, {Coppa}, {Cucciati}, {de la Torre},
  {de Ravel}, {Franzetti}, {Garilli}, {Hasinger}, {Iovino}, {Kampczyk},
  {Kneib}, {Knobel}, {Kovac}, {Le Borgne}, {Le Brun}, {F{\`e}vre}, {Lilly},
  {Looper}, {Maier}, {Mainieri}, {Mellier}, {Mignoli}, {Murayama}, {Pell{\`o}},
  {Peng}, {P{\'e}rez-Montero}, {Renzini}, {Ricciardelli}, {Schiminovich},
  {Scodeggio}, {Shioya}, {Silverman}, {Surace}, {Tanaka}, {Tasca}, {Tresse},
  {Vergani}, \& {Zucca}}]{ilbert_2009}
---. 2009, \apj, 690, 1236

\bibitem[{{Jouvel} {et~al.}(2009){Jouvel}, {Kneib}, {Ilbert}, {Bernstein},
  {Arnouts}, {Dahlen}, {Ealet}, {Milliard}, {Aussel}, {Capak}, {Koekemoer}, {Le
  Brun}, {McCracken}, {Salvato}, \& {Scoville}}]{jouvel_2009}
{Jouvel}, S., {et~al.} 2009, \aap, 504, 359

\bibitem[{{Kennicutt}(1998)}]{kennicutt_1998}
{Kennicutt}, Jr., R.~C. 1998, \araa, 36, 189

\bibitem[{{King}(2005)}]{king_2005}
{King}, L.~J. 2005, \aap, 441, 47

\bibitem[{{Kirk} {et~al.}(2010){Kirk}, {Bridle}, \& {Schneider}}]{kirk_2010}
{Kirk}, D., {Bridle}, S., \& {Schneider}, M. 2010, \mnras, 408, 1502

\bibitem[{{Kitching} {et~al.}(2007){Kitching}, {Heavens}, {Taylor}, {Brown},
  {Meisenheimer}, {Wolf}, {Gray}, \& {Bacon}}]{kitching_2007}
{Kitching}, T.~D., {Heavens}, A.~F., {Taylor}, A.~N., {Brown}, M.~L.,
  {Meisenheimer}, K., {Wolf}, C., {Gray}, M.~E., \& {Bacon}, D.~J. 2007,
  \mnras, 376, 771

\bibitem[{{Komatsu} {et~al.}(2011){Komatsu}, {Smith}, {Dunkley}, {Bennett},
  {Gold}, {Hinshaw}, {Jarosik}, {Larson}, {Nolta}, {Page}, {Spergel},
  {Halpern}, {Hill}, {Kogut}, {Limon}, {Meyer}, {Odegard}, {Tucker}, {Weiland},
  {Wollack}, \& {Wright}}]{komatsu_2011}
{Komatsu}, E., {et~al.} 2011, \apjs, 192, 18

\bibitem[{{Lesser} \& {Tyson}(2002)}]{lesser_2002}
{Lesser}, M.~P., \& {Tyson}, J.~A. 2002, in Society of Photo-Optical
  Instrumentation Engineers (SPIE) Conference Series, Vol. 4836, Society of
  Photo-Optical Instrumentation Engineers (SPIE) Conference Series, ed.
  {J.~A.~Tyson \& S.~Wolff}, 240--246

\bibitem[{{LSST Science Collaboration}(2009)}]{lsst_2009}
{LSST Science Collaboration}. 2009, ArXiv e-prints

\bibitem[{{Ly} {et~al.}(2011){Ly}, {Malkan}, {Hayashi}, {Motohara},
  {Kashikawa}, {Shimasaku}, {Nagao}, \& {Grady}}]{ly_2011}
{Ly}, C., {Malkan}, M.~A., {Hayashi}, M., {Motohara}, K., {Kashikawa}, N.,
  {Shimasaku}, K., {Nagao}, T., \& {Grady}, C. 2011, \apj, 735, 91

\bibitem[{{MacDonald} \& {Bernstein}(2010)}]{macdonald_2010}
{MacDonald}, C.~J., \& {Bernstein}, G. 2010, \pasp, 122, 485

\bibitem[{{Madau}(1995)}]{madau_1995}
{Madau}, P. 1995, \apj, 441, 18

\bibitem[{{Mandelbaum} {et~al.}(2006){Mandelbaum}, {Hirata}, {Broderick},
  {Seljak}, \& {Brinkmann}}]{mandelbaum_2006}
{Mandelbaum}, R., {Hirata}, C.~M., {Broderick}, T., {Seljak}, U., \&
  {Brinkmann}, J. 2006, \mnras, 370, 1008

\bibitem[{{Mandelbaum} {et~al.}(2011){Mandelbaum}, {Blake}, {Bridle},
  {Abdalla}, {Brough}, {Colless}, {Couch}, {Croom}, {Davis}, {Drinkwater},
  {Forster}, {Glazebrook}, {Jelliffe}, {Jurek}, {Li}, {Madore}, {Martin},
  {Pimbblet}, {Poole}, {Pracy}, {Sharp}, {Wisnioski}, {Woods}, \&
  {Wyder}}]{mandelbaum_2011}
{Mandelbaum}, R., {et~al.} 2011, \mnras, 410, 844

\bibitem[{{Mazin} {et~al.}(2012){Mazin}, {Bumble}, {Meeker}, {O'Brien},
  {McHugh}, \& {Langman}}]{mazin_2012}
{Mazin}, B.~A., {Bumble}, B., {Meeker}, S.~R., {O'Brien}, K., {McHugh}, S., \&
  {Langman}, E. 2012, Optics Express, 20, 1503

\bibitem[{{Mazin} {et~al.}(2010){Mazin}, {O'Brien}, {McHugh}, {Bumble},
  {Moore}, {Golwala}, \& {Zmuidzinas}}]{mazin_2010}
{Mazin}, B.~A., {O'Brien}, K., {McHugh}, S., {Bumble}, B., {Moore}, D.,
  {Golwala}, S., \& {Zmuidzinas}, J. 2010, in Society of Photo-Optical
  Instrumentation Engineers (SPIE) Conference Series, Vol. 7735, Society of
  Photo-Optical Instrumentation Engineers (SPIE) Conference Series

\bibitem[{{Mazin} {et~al.}(2013){Mazin}, {Meeker}, {Strader}, {Bumble},
  {O'Brien}, {Szypryt}, {Marsden}, {van Eyken}, {Duggan}, {Ulbricht},
  {Stoughton}, \& {Johnson}}]{mazin_2013}
{Mazin}, B.~A., {et~al.} 2013, ArXiv e-prints

\bibitem[{{McHugh} {et~al.}(2012){McHugh}, {Mazin}, {Serfass}, {Meeker},
  {O'Brien}, {Duan}, {Raffanti}, \& {Werthimer}}]{mchugh_2012}
{McHugh}, S., {Mazin}, B.~A., {Serfass}, B., {Meeker}, S., {O'Brien}, K.,
  {Duan}, R., {Raffanti}, R., \& {Werthimer}, D. 2012, Review of Scientific
  Instruments, 83, 044702

\bibitem[{{Moles} {et~al.}(2008){Moles}, {Ben{\'{\i}}tez}, {Aguerri}, {Alfaro},
  {Broadhurst}, {Cabrera-Ca{\~n}o}, {Castander}, {Cepa}, {Cervi{\~n}o},
  {Crist{\'o}bal-Hornillos}, {Fern{\'a}ndez-Soto}, {Gonz{\'a}lez Delgado},
  {Infante}, {M{\'a}rquez}, {Mart{\'{\i}}nez}, {Masegosa}, {del Olmo}, {Perea},
  {Prada}, {Quintana}, \& {S{\'a}nchez}}]{moles_2008}
{Moles}, M., {et~al.} 2008, \aj, 136, 1325

\bibitem[{{Newman}(2008)}]{newman_2008}
{Newman}, J.~A. 2008, \apj, 684, 88

\bibitem[{{O'Brien} {et~al.}(2012){O'Brien}, {Mazin}, {McHugh}, {Meeker}, \&
  {Bumble}}]{obrien_2012}
{O'Brien}, K., {Mazin}, B., {McHugh}, S., {Meeker}, S., \& {Bumble}, B. 2012,
  in IAU Symposium, Vol. 285, IAU Symposium, 385--388

\bibitem[{{Padmanabhan} {et~al.}(2012){Padmanabhan}, {Xu}, {Eisenstein},
  {Scalzo}, {Cuesta}, {Mehta}, \& {Kazin}}]{padmanabhan_2012}
{Padmanabhan}, N., {Xu}, X., {Eisenstein}, D.~J., {Scalzo}, R., {Cuesta},
  A.~J., {Mehta}, K.~T., \& {Kazin}, E. 2012, ArXiv e-prints

\bibitem[{{Padmanabhan} {et~al.}(2007){Padmanabhan}, {Schlegel}, {Seljak},
  {Makarov}, {Bahcall}, {Blanton}, {Brinkmann}, {Eisenstein}, {Finkbeiner},
  {Gunn}, {Hogg}, {Ivezi{\'c}}, {Knapp}, {Loveday}, {Lupton}, {Nichol},
  {Schneider}, {Strauss}, {Tegmark}, \& {York}}]{padmanabhan_2007}
{Padmanabhan}, N., {et~al.} 2007, \mnras, 378, 852

\bibitem[{{Peacock} {et~al.}(2006){Peacock}, {Schneider}, {Efstathiou},
  {Ellis}, {Leibundgut}, {Lilly}, \& {Mellier}}]{peacock_2006}
{Peacock}, J.~A., {Schneider}, P., {Efstathiou}, G., {Ellis}, J.~R.,
  {Leibundgut}, B., {Lilly}, S.~J., \& {Mellier}, Y. 2006, {ESA-ESO Working
  Group on ''Fundamental Cosmology''}, Tech. rep.

\bibitem[{{Perlmutter} {et~al.}(1999){Perlmutter}, {Aldering}, {Goldhaber},
  {Knop}, {Nugent}, {Castro}, {Deustua}, {Fabbro}, {Goobar}, {Groom}, {Hook},
  {Kim}, {Kim}, {Lee}, {Nunes}, {Pain}, {Pennypacker}, {Quimby}, {Lidman},
  {Ellis}, {Irwin}, {McMahon}, {Ruiz-Lapuente}, {Walton}, {Schaefer}, {Boyle},
  {Filippenko}, {Matheson}, {Fruchter}, {Panagia}, {Newberg}, {Couch}, \&
  {Supernova Cosmology Project}}]{perlmutter_1999}
{Perlmutter}, S., {et~al.} 1999, \apj, 517, 565

\bibitem[{{Richards} {et~al.}(2009){Richards}, {Myers}, {Gray}, {Riegel},
  {Nichol}, {Brunner}, {Szalay}, {Schneider}, \& {Anderson}}]{richards_2009}
{Richards}, G.~T., {et~al.} 2009, \apjs, 180, 67

\bibitem[{{Riess} {et~al.}(1998){Riess}, {Filippenko}, {Challis},
  {Clocchiatti}, {Diercks}, {Garnavich}, {Gilliland}, {Hogan}, {Jha},
  {Kirshner}, {Leibundgut}, {Phillips}, {Reiss}, {Schmidt}, {Schommer},
  {Smith}, {Spyromilio}, {Stubbs}, {Suntzeff}, \& {Tonry}}]{reiss_1998}
{Riess}, A.~G., {et~al.} 1998, \aj, 116, 1009

\bibitem[{{Roesch} {et~al.}(2010){Roesch}, {Bideaud}, {Benoit}, {Cruciani},
  {D{\'e}sert}, {Doyle}, {Leclercq}, {Mattiocco}, {Schuster}, {Swenson}, \&
  {Monfardini}}]{roesch_2010}
{Roesch}, M., {et~al.} 2010, in Society of Photo-Optical Instrumentation
  Engineers (SPIE) Conference Series, Vol. 7741, Society of Photo-Optical
  Instrumentation Engineers (SPIE) Conference Series

\bibitem[{{Sawangwit} {et~al.}(2012){Sawangwit}, {Shanks}, {Croom},
  {Drinkwater}, {Fine}, {Parkinson}, \& {Ross}}]{sawangwit_2012}
{Sawangwit}, U., {Shanks}, T., {Croom}, S.~M., {Drinkwater}, M.~J., {Fine}, S.,
  {Parkinson}, D., \& {Ross}, N.~P. 2012, \mnras, 420, 1916

\bibitem[{{Schlaerth} {et~al.}(2010){Schlaerth}, {Czakon}, {Day}, {Downes},
  {Duan}, {Gao}, {Glenn}, {Golwala}, {Hollister}, {Leduc}, {Mazin}, {Maloney},
  {Noroozian}, {Nguyen}, {Sayers}, {Siegel}, {Vaillancourt}, {Vayonakis},
  {Wilson}, \& {Zmuidzinas}}]{schlaerth_2010}
{Schlaerth}, J.~A., {et~al.} 2010, in Society of Photo-Optical Instrumentation
  Engineers (SPIE) Conference Series, Vol. 7741, Society of Photo-Optical
  Instrumentation Engineers (SPIE) Conference Series

\bibitem[{{Schlegel} {et~al.}(2009){Schlegel}, {White}, \&
  {Eisenstein}}]{schlegel_2009}
{Schlegel}, D., {White}, M., \& {Eisenstein}, D. 2009, in ArXiv Astrophysics
  e-prints, Vol. 2010, astro2010: The Astronomy and Astrophysics Decadal
  Survey, 314

\bibitem[{{Schlegel} {et~al.}(2011){Schlegel}, {Abdalla}, {Abraham}, {Ahn},
  {Allende Prieto}, {Annis}, {Aubourg}, {Azzaro}, {Baltay}, {Baugh}, {Bebek},
  {Becerril}, {Blanton}, {Bolton}, {Bromley}, {Cahn}, {Carton},
  {Cervantes-Cota}, {Chu}, {Cortes}, {Dawson}, {Dey}, {Dickinson}, {Diehl},
  {Doel}, {Ealet}, {Edelstein}, {Eppelle}, {Escoffier}, {Evrard}, {Faccioli},
  {Frenk}, {Geha}, {Gerdes}, {Gondolo}, {Gonzalez-Arroyo}, {Grossan},
  {Heckman}, {Heetderks}, {Ho}, {Honscheid}, {Huterer}, {Ilbert}, {Ivans},
  {Jelinsky}, {Jing}, {Joyce}, {Kennedy}, {Kent}, {Kieda}, {Kim}, {Kim},
  {Kneib}, {Kong}, {Kosowsky}, {Krishnan}, {Lahav}, {Lampton}, {LeBohec}, {Le
  Brun}, {Levi}, {Li}, {Liang}, {Lim}, {Lin}, {Linder}, {Lorenzon}, {de la
  Macorra}, {Magneville}, {Malina}, {Marinoni}, {Martinez}, {Majewski},
  {Matheson}, {McCloskey}, {McDonald}, {McKay}, {McMahon}, {Menard},
  {Miralda-Escude}, {Modjaz}, {Montero-Dorta}, {Morales}, {Mostek}, {Newman},
  {Nichol}, {Nugent}, {Olsen}, {Padmanabhan}, {Palanque-Delabrouille}, {Park},
  {Peacock}, {Percival}, {Perlmutter}, {Peroux}, {Petitjean}, {Prada},
  {Prieto}, {Prochaska}, {Reil}, {Rockosi}, {Roe}, {Rollinde}, {Roodman},
  {Ross}, {Rudnick}, {Ruhlmann-Kleider}, {Sanchez}, {Sawyer}, {Schimd},
  {Schubnell}, {Scoccimaro}, {Seljak}, {Seo}, {Sheldon}, {Sholl},
  {Shulte-Ladbeck}, {Slosar}, {Smith}, {Smoot}, {Springer}, {Stril}, {Szalay},
  {Tao}, {Tarle}, {Taylor}, {Tilquin}, {Tinker}, {Valdes}, {Wang}, {Wang},
  {Weaver}, {Weinberg}, {White}, {Wood-Vasey}, {Yang}, {Yeche}, {Zakamska},
  {Zentner}, {Zhai}, \& {Zhang}}]{schlegel_2011}
{Schlegel}, D., {et~al.} 2011, ArXiv e-prints

\bibitem[{{Seo} {et~al.}(2011){Seo}, {Sato}, {Dodelson}, {Jain}, \&
  {Takada}}]{seo_2011}
{Seo}, H.-J., {Sato}, M., {Dodelson}, S., {Jain}, B., \& {Takada}, M. 2011,
  \apjl, 729, L11

\bibitem[{{Silk}(1968)}]{silk_1968}
{Silk}, J. 1968, \apj, 151, 459

\bibitem[{{Smail} {et~al.}(1995){Smail}, {Hogg}, {Yan}, \&
  {Cohen}}]{smail_1995}
{Smail}, I., {Hogg}, D.~W., {Yan}, L., \& {Cohen}, J.~G. 1995, \apjl, 449, L105

\bibitem[{{Springel} {et~al.}(2005){Springel}, {White}, {Jenkins}, {Frenk},
  {Yoshida}, {Gao}, {Navarro}, {Thacker}, {Croton}, {Helly}, {Peacock}, {Cole},
  {Thomas}, {Couchman}, {Evrard}, {Colberg}, \& {Pearce}}]{springel_2005}
{Springel}, V., {et~al.} 2005, \nat, 435, 629

\bibitem[{{Takada} \& {White}(2004)}]{takada_2004}
{Takada}, M., \& {White}, M. 2004, \apjl, 601, L1

\bibitem[{{The Dark Energy Survey Collaboration}(2005)}]{des_2005}
{The Dark Energy Survey Collaboration}. 2005, ArXiv Astrophysics e-prints

\bibitem[{{Vanden Berk} {et~al.}(2001){Vanden Berk}, {Richards}, {Bauer},
  {Strauss}, {Schneider}, {Heckman}, {York}, {Hall}, {Fan}, {Knapp},
  {Anderson}, {Annis}, {Bahcall}, {Bernardi}, {Briggs}, {Brinkmann}, {Brunner},
  {Burles}, {Carey}, {Castander}, {Connolly}, {Crocker}, {Csabai}, {Doi},
  {Finkbeiner}, {Friedman}, {Frieman}, {Fukugita}, {Gunn}, {Hennessy},
  {Ivezi{\'c}}, {Kent}, {Kunszt}, {Lamb}, {Leger}, {Long}, {Loveday}, {Lupton},
  {Meiksin}, {Merelli}, {Munn}, {Newberg}, {Newcomb}, {Nichol}, {Owen}, {Pier},
  {Pope}, {Rockosi}, {Schlegel}, {Siegmund}, {Smee}, {Snir}, {Stoughton},
  {Stubbs}, {SubbaRao}, {Szalay}, {Szokoly}, {Tremonti}, {Uomoto}, {Waddell},
  {Yanny}, \& {Zheng}}]{vandenberk_2001}
{Vanden Berk}, D.~E., {et~al.} 2001, \aj, 122, 549

\bibitem[{{Wang} {et~al.}(2010){Wang}, {Percival}, {Cimatti}, {Mukherjee},
  {Guzzo}, {Baugh}, {Carbone}, {Franzetti}, {Garilli}, {Geach}, {Lacey},
  {Majerotto}, {Orsi}, {Rosati}, {Samushia}, \& {Zamorani}}]{wang_2010}
{Wang}, Y., {et~al.} 2010, \mnras, 409, 737

\bibitem[{{Weinberg} {et~al.}(2012){Weinberg}, {Mortonson}, {Eisenstein},
  {Hirata}, {Riess}, \& {Rozo}}]{weinberg_2012}
{Weinberg}, D.~H., {Mortonson}, M.~J., {Eisenstein}, D.~J., {Hirata}, C.,
  {Riess}, A.~G., \& {Rozo}, E. 2012, ArXiv e-prints

\bibitem[{{Wolf} {et~al.}(2004){Wolf}, {Meisenheimer}, {Kleinheinrich},
  {Borch}, {Dye}, {Gray}, {Wisotzki}, {Bell}, {Rix}, {Cimatti}, {Hasinger}, \&
  {Szokoly}}]{wolf_2004}
{Wolf}, C., {et~al.} 2004, \aap, 421, 913

\end{thebibliography}
\end{document}